\documentstyle[aaspp4]{article}
\def\ea{{\it et al.} }
\def\ref{\par\noindent\hangindent 20 pt}

\def\sec{^{\prime\prime\!\!}}
\def\dvac{de~Vaucouleurs } 
\def\wsy{Wurtz, Stocke \& Yee (1996)}

\def\ntot{110 } 

\def\nres{72 }  

\def\resolved{69 } 

\def\cls{58 }   

\def\ucls{14 }  

\def\low_z{63 } 

\def\res_low_z{58 } 

\def\high_z{23 } 

\def\res_high_z{6 } 

\begin{document}
\input{psfig.tex}


\title{ The $HST$ Survey of BL Lacertae Objects. I. Surface 
Brightness Profiles, Magnitudes, and Radii of Host Galaxies }

\author{Riccardo Scarpa and C. Megan Urry} 
\authoraddr{Space Telescope Science Institute, 3700 San Martin Dr., 
Baltimore, MD 21218, USA}
\authoremail{scarpa@stsci.edu,cmu@stsci.edu}
\affil{Space Telescope Science Institute}

\author{Renato Falomo} 
\authoraddr{Osservatorio Astronomico di Padova, Vicolo dell'Osservatorio 5, 
35122 Padova, Italy}
\authoremail{falomo@astrpd.pd.astro.it}
\affil{Astronomical Observatory of Padua}

\author{Joseph E. Pesce} 
\authoraddr{Department of Astronomy, Pennsylvania State University,
525 Davey Lab, University Park, PA 16802 }
\affil{Eureka Scientific, Inc.}
\authoremail{pesce@astro.psu.edu}

\author{Aldo Treves} 
\authoraddr{Como, Italy}
\authoremail{treves@uni.mi.astro.it}
\affil{University of Insubria}

\begin{abstract}

We report on a large $HST$ imaging survey of BL Lac objects,
at spatial resolution $\gtrsim 10$ times better than previous 
ground-based surveys. 
We focus on data reduction and analysis, describing the procedures used to
model the host galaxy surface brightness radial profiles. 
A total of \resolved\ host galaxies were resolved out of \ntot\
objects observed, including almost all sources at $z\lesssim 0.5$. We
classify them morphologically by fitting with either an exponential
disk or a de~Vaucouleurs profile; when one fit is preferred over the
other, in \cls\ of \resolved\ cases, it is invariably the elliptical
morphology. This is a very strong result given the large number of BL
Lac objects, the unprecedented spatial resolution, and the homogeneity
of the data set. With the present reclassification of the host galaxy
of 1418+546 as an elliptical, there remain no undisputed examples of a
disk galaxy hosting a BL Lac nucleus. This implies that, at 99\%
confidence, fewer than 7\% of BL Lacs can be in disk galaxies. The
apparent magnitude of the host galaxies varies with distance as
expected if the absolute magnitudes are approximately the same, with a
spread of $\pm 1$~mag, out to redshift $z\sim0.5$. At larger
redshifts, only \res_high_z\ of \high_z\ BL Lacs are resolved so the
present data do not constrain possible luminosity evolution of the
host galaxies. The collective Hubble diagram for BL Lac host galaxies
and radio galaxies strongly supports their unification.

Subject headings: BL Lacertae objects ---
galaxies: structure --- galaxies: elliptical

\end{abstract}

\section{Introduction}

Determining the properties of AGN host galaxies is an important approach
to understanding the AGN phenomenon. The Hubble Space
Telescope ($HST$\footnote{Based on observations made with the NASA/ESA
Hubble Space Telescope, obtained at the Space Telescope Science 
Institute, which is operated by the Association of Universities 
for Research in Astronomy, Inc., under NASA contract NAS~5-26555.}),
because of its superb spatial resolution, is a valuable tool for this
kind of investigation, and indeed this was one of the key science 
objectives of the $HST$.
Not surprisingly, $HST$ has been used extensively to image AGN 
host galaxies (Disney \ea 1995; Bahcall \ea 1997; Hooper, Impey
\& Fultz 1997; Falomo \ea 1997; Malkan, Gorjian \& Raymond 1998; Urry
\ea 1999; McLeod, Rieke \& Storrie-Lombardi 1999; McLure \ea 1999).

In the current paradigm for radio-loud AGN,
BL Lac objects have relativistically out-flowing 
jets oriented nearly along the line of sight (Urry \& Padovani 1995).
Strong relativistic beaming of the jet emission then alters the
observed properties of these blazars. 
Radio-loud AGN pointing at different angles are seen as quasars or radio
galaxies --- this is the so-called ``unification'' picture.
Proving the correctness of unified schemes
is of major importance for our understanding of the true nature
of AGN, and is among the most lively topics in modern
astrophysics.

Based on preliminary surveys and host galaxy properties, 
BL Lacs were identified early on with low-luminosity radio galaxies
(Schwartz \& Ku 1983, Perez-Fournon \& Biermann 1984, Ulrich 1989, Browne 1989),
i.e., Fanaroff \& Riley type I (FR~I) radio galaxies (Fanaroff \& Riley 1974). 
Many subsequent studies have supported this hypothesis
(Urry \& Padovani 1995, and references therein). However, as is
often the case in nature, the picture is not so simple and some
authors have proposed more complex scenarios in which the
parent population includes only some FR~Is (e.g., Wurtz, Stocke
\& Yee 1996), or a mix of FR~I and FR~II (Kollgaard \ea 1996). Indeed, 
the extended radio morphologies of BL Lacs 
can be of both FR~I and II types (Kollgaard \ea 1992), and some of 
the line emission from FR~IIs is weak enough to be BL Lac-like 
(Laing \ea 1994). Thus it may be appropriate to unify BL Lacs more
generally with radio galaxies.

The luminosity, size, morphology, and global structure of the BL Lac
host galaxy are clearly unaffected by beaming, and so can be a
powerful tool to test unified schemes. With ground-based data,
several authors have investigated this subject, with consistent
results. The BL Lac host galaxies are almost always giant ellipticals,
$\sim 1$~mag brighter than an $M^*$ galaxy, with effective radius of
several kiloparsecs (Abraham, McHardy \& Crawford 1991; Stickel, Fried
\& K\"uhr 1993; Wurtz, Stocke \& Yee 1996; Falomo 1996). Similar
results have also been found for a handful of BL Lacs observed with
$HST$ (Falomo \ea 1997; Jannuzi, Yanny \& Impey 1997; Urry \ea 1999).
The general properties of the BL Lacs hosts appear consistent with the
prediction of the unified models, since the optical counterparts of
radio galaxies are always giant ellipticals. However, differences do
exist when the average properties of different samples are
compared. In particular it is not yet clear whether BL Lac host
galaxies are more similar to FR~I or to FR~II hosts. The issue
therefore remains open and is further complicated by claims that some
BL Lacs reside in disk galaxies (Halpern \ea 1986; Abraham, McHardy \&
Crawford 1991; Wurtz, Stocke \& Yee 1996).

To further investigate this issue, we carried out a large $HST$
imaging survey of BL Lac objects, observing \ntot\ objects in
``snapshot'' mode starting from 132 BL Lacs from seven complete
flux-limited samples spanning the redshift range $0.031<z<1.34$. Here
we describe the data reduction and analysis procedures in some detail,
and present the $HST$ images, the profile fits, and the associated
chi-squared contours. Properties of the host galaxies are discussed
in more detail by Urry \ea 2000 (hereafter Paper~II). This paper is
organized as follows. In \S~2 we describe observations and data
analysis. \S~ 3 reports results, which are more fully described in
Paper~II, and compares them with previous ground-based surveys.
Conclusions are in \S~4. Details of individual sources are given in
the Appendix, and some of the more peculiar sources, as well as the
discovery of an optical jet in PKS~2201+044, have been presented
previously by Scarpa \ea (1999a,1999b).

\section{Observations and Data Analysis}

Observations were carried out as $HST$ snapshots, which are short
exposures obtained during gaps in the observing schedule. A total of
\ntot\ targets were selected at random (i.e., depending only on
details of the gaps) from an initial sample of 132 BL Lacs from seven
flux-limited samples. These were observed with the Wide Field and
Planetary Camera 2 (WFPC2) through the F702W, or in few cases F606W or
F555W, filter. For optimal point spread function (PSF) sampling and 
stability, all objects were placed near the center of the PC chip.

The journal of the observations is reported in Table~1. It includes
for each source the original flux-limited sample from which it came,
the object type (high- or low-frequency peaked), redshift, and
position. Table~1 also lists the date of the $HST$ observation,
exposure time, filter, and measured surface-brightness of the sky
background for that image.

To obtain for each target a final image well exposed both in the
inner, bright nucleus and in the outer regions where the host galaxy
emission should be well above the wings of the point spread function,
we set up a series of exposures with duration ranging from a few tens
to up $\sim 300$ seconds. For each object we usually obtained 3-5
images, which were later combined, to remove cosmic-ray events and
improve signal-to-noise ratio. This was done 
with the CRREJ task available within IRAF, which first masks all
saturated pixels, then rescales each image to a common exposure time 
in order to look for deviant pixels (cosmic rays), and finally adds
the images (compensating for pixels eventually masked).
The final combined images are shown in
Figure~1.

\subsection{Flux Calibration}

Images were flux calibrated following the prescription of
Holtzman \ea (1995, their Eq.~9 and Table~10). Namely, we convert
F702W and F606W fluxes to Cousin R and Johnson V
magnitudes, respectively, with the following transformations:

\begin{displaymath}
 m_R = -2.5 \log({\rm DN_{F702W}}) + 21.511 +0.486 (V - R) -0.079 (V -
R)^2 + 2.5 \log({\rm GN}) + 0.1 ,
\end{displaymath}
\begin{displaymath}
 m_V = -2.5 \log({\rm DN_{F606W}}) + 22.093 +0.254 (V - I) +0.012 (V -
I)^2 + 2.5 \log({\rm GN}) + 0.1 ~,
\end{displaymath}

\noindent
where DN is the digit number per second, GN=2 in the case of gain=7
(our case), and the constant term 0.1~mag corrects for our use of an infinite
aperture compared to Holtzman's 0.5-arcsec aperture. 
For the sake of clarity, the magnitudes of the few objects 
observed in filters other than F702W were transformed into
the R band; we assumed $V-R=0.3$ for the point source, and for the host the
appropriate $V-R$ color for an elliptical galaxy at the 
redshift of the BL Lac, as reported in Table~3. 

\subsection{$HST$ Point Spread Function}

For studying AGN host galaxies, knowledge of the shape and stability
of the PSF is of major importance. 
We carefully developed a PSF model and tested its reliability extensively.
As a first step, to minimize the effects of the spatial 
variation of the PSF, 
we placed our targets always at the center of the PC field of
view (within 50 pixels).
Then we used the theoretical model produced by the Tiny
Tim software (Krist 1995), which gives an excellent two-dimensional
representation of the PSF within $\sim 2$~arcsec. Outside this range,
particularly in the PC camera, there is a substantial contribution due
to large-angle scattered light, which is not included in the Tiny Tim model.
Indeed, we found that the radial profiles of several high redshift, 
unresolved BL Lac objects lay systematically above the Tiny Tim PSF,
simply due to scattered light. 

To construct the wings of the PSF, we took from the $HST$ archive 
several images of isolated and extremely over-exposed stars.
These stars were saturated to such an extent that the PSF wings 
spread over the three WF chips as well as the PC.
Comparing the radial profiles of the different stars,
we found that the fraction of scattered light is roughly constant,
and only marginally sensitive to telescope focus and position on the chip.
We built our final PSF model by smoothly joining, at 2~arcsec,
the Tiny Tim inner profile to the average profile of the stars. 
Figure~2 shows the final composite PSF and individual data
points for the unsaturated parts of the stellar images.
This composite profile fits very well out to $\sim 7$~arcsec 
from the center, spanning a range of more than 15 magnitudes 
(a factor $10^6$ in flux).
Small and unpredictable variations of the PSF due to
telescope ``breathing'' were not modeled but were taken into account
during the fitting procedure (see \S~2.3).
For the few sources not 
at the center of the PC (see notes to the objects), 
we built a PSF model appropriate to each position.

We give here a simple analytical formula for describing the fraction of 
scattered light as a function of the total flux $I_{PSF}$ of the Tiny 
Tim model. On the one dimensional radial profile, the excess light $S$ to 
be added to the Tiny Tim model at distance $r$~arcsec from the center is:
\begin{displaymath}
S(r) = A\ I_{PSF}\ e^{B\ r}.
\end{displaymath}
The constants $A$ and $B$ depend weakly on wavelength, and are listed in
Table~2 for all four filters considered in this paper.
The parameter $B$ is quite important as it determines the slope
of the scattered light profile, and hence the final 
slope of the PSF wing. 

Properly taking into account the scattered light is of course important 
in determining whether an object is resolved, but it is also
very important when the surface brightness of the host galaxy is
comparable to that of the PSF wings. If not included in the PSF
model, the scattered light will be attributed to the host, significantly
overestimating its luminosity, or leading to spurious galaxy
detections. Moreover, a spurious correlation between host and point
source luminosity would be introduced, because scattered light and point
source luminosity are proportional.
 
\subsection{One-Dimensional Profile and Fitting Procedure}

Our snapshot images are typically shorter than would have been ideal
for studying extended emission from the host galaxy. As a result, a
full two-dimensional analysis of the data has been done by Falomo \ea
(2000) only for the $\sim30$ low redshift sources, for which the host
galaxy is well exposed. Instead, we consider here only the
azimuthally averaged surface brightness radial profile, thereby
increasing the signal-to-noise ratio at the expense of spatial
information. In particular, information about asymmetries and/or
non-circularities of the host galaxy is lost in the one-dimensional
approach. 

The median surface brightness reached in our data is
$\mu_R=25.2$ mag/arcsec$^2$, 2.5 magnitudes fainter than
the sky median surface brightness, allowing us to trace a typical
host galaxy out four effective radii. The 1-dimensional
radial profiles were determined by averaging the flux over annuli
spaced at 1-pixel intervals. Nearby companions and/or stars were masked out
and that region of the image simply not considered.

The statistical error associated with the
source flux in each annulus was computed considering the statistical
noise, readout noise, sky fluctuations, and digitization noise. Sky
flux was estimated averaging over several regions
uniformly distributed around the chip. The radial surface
brightness profiles of all \ntot\ sources is shown in Figure~1.

Radial profiles were fitted with models consisting of PSF plus galaxy;
the latter was modeled with either a \dvac $r^{1/4}$ law or an exponential 
law convolved with the PSF. Rather than subtracting the PSF
first and analyzing the residuals (the galaxy light), we allowed
the PSF normalization and the galaxy brightness to vary simultaneously
and independently, minimizing the $\chi^2$ to find the best fit.
This avoids the {\it a priori} determination of the PSF
normalization, which can introduce systematic bias in the derived
galaxy luminosities. 

The fit involves three free parameters varied simultaneously: 
the PSF total flux, the host
galaxy total flux, and the host galaxy effective radius. The best-fit
values were then determined from $\chi^2$ minimization.
Several details are important to note.
First, because the PSF is under-sampled, we excluded the central
0.1~arcsec (the first 3 pixels) from the fit. Second, to account for
possible PSF variations introduced by telescope ``breathing'', a
systematic uncertainty in the PSF model was included. This error was
assumed to be 10\% of the point source contribution in each pixel, and was
added in quadrature to the errors originally associated to the radial
profile. In order to actually compute this extra source of error, we
need to know at least roughly the PSF normalization. So we first did
a coarse fit to normalize the PSF sufficiently well to
fix the systematic errors (not to fix the PSF), which are only
approximation in any case. We then did the actual fitting for the best-fit
PSF and galaxy. The $\chi^2$ distribution was used to
evaluate the statistical uncertainties on the three fit parameters.

To determine whether an object was resolved, we fitted the
radial profile with a PSF only and with a PSF plus either an
exponential disk or a \dvac $r^{1/4}$ law; we then compared, via an
F-test, the best-fit $\chi^2$ values for the 3 cases. 
Our threshold was that a galaxy was formally detected when one of the 
galaxy models was preferred over the PSF only at $>99$\% confidence. 
To be conservative, and to avoid being fooled by large-scale fluctuations
in the sky background, we further required that after increasing the
sky flux by $1\sigma$, the addition of a galaxy still improved the fit 
at $>90$\% confidence.
Similarly, we evaluated which of the two galaxy
models was preferred using an F-test at the 99\% confidence level.

Fits to the radial profiles are shown in Figure~1 and quantitative
results from the one-dimensional fitting are given in Table~3. In
addition to the best-fit parameters for \dvac and disk models, Table~3
lists a model-independent total apparent magnitude (column 5),
integrated to the last point of the radial profiles shown in Figure~1,
as well as the chi-squared for a point-source-only fit and the number
of data points in the radial profile (column 6). In general, the
best-fit PSF normalizations (i.e., nuclear magnitudes) differ for the
\dvac and disk models, the value associated with the disk model being
systematically brighter since it compensates for a relative lack of
light in the galaxy center. Only when we can discriminate between the
two galaxy models do we have an unambiguous estimate of the point
source flux.

We assessed statistical uncertainties in the fit parameters from
multi-dimensional $\chi^2$ confidence contours, which are also 
shown in Figure~1 (for resolved objects).
Quoted in Table~3 are the 68\% ($1\sigma$) confidence
uncertainties ($\Delta \chi^2 = 2.3$ for the two parameters of
interest, host galaxy magnitude and half-light radius) from the box
circumscribing the contour. These represent statistical uncertainties
depends only on our estimate of the error bars, and is not related with 
the fitting procedure. They are in most cases very small, but these 
smallness should not give the impression that 
the quantities are actually determined so precisely, because the effect of 
systematic errors in not included in these values.
Indeed, comparing the results obtained from different groups starting
from the same set of HST data (e.g., Urry \ea 1999), we know that the usual 
difference in the
final estimated host magnitude is larger, some times $\gtrsim
0.2$~mag. In several cases this is much larger than the estimated
statistical uncertainty, indicating that systematic errors are dominant. 

Finally, for the unresolved BL~Lac objects we determined 99\%
confidence upper limits (statistical errors) to the host galaxy
magnitudes ($\Delta \chi^2 = 6.6$ for one parameter of interest,
$M_{gal}$), fixing the effective radius at $r_e =10$~kpc, slightly larger
than the median (to be conservative) for the resolved objects.

\subsection{Simulations and Systematic Errors}

To test the reliability of our results we performed a series of 140
simulations. We created simulated data by combining a central point
source (represented by a two-dimensional composite PSF template) with
an elliptical host galaxy ranging in brightness from the point source
magnitude to 3 magnitudes fainter. All relevant parameters were
adjusted to obtain similar counts per pixel as in the real images, and
statistical noise was added. Analysis of the radial profile then
followed exactly the same procedure adopted for real objects.

We recover the input parameters with high precision 
and no systematic biases. The three histograms in Figure~3 show,
for each of the three fit parameters (point source magnitude, 
host galaxy magnitude, and effective radius)
the difference between the input parameter value used to generate the
simulated data and the fitted parameter derived by minimizing $\chi^2$.
The expected values were recovered to within $\sim 10$\%,
with a narrow peak centered at the expected value. 
The effective radius is the least precisely determined of the three
parameters, being off by as much as 0.5~arcsec in our simulations (still
much less than disagreement with ground-based measurements; see \S~3.2).
Most important, no systematic deviations from the expected values are
observed, demonstrating that our analysis procedures do not induce spurious
trends in the final results.
Similar results were obtained when fitting simulated PSF plus disk galaxies,
with the effective radius being again the less precisely determined parameter.

\section{Results}

\subsection{Detection Rate of Host Galaxies }

Figure~1 shows for each source a gray-scale image representing the
central part of the PC camera, together with an isophotal contour
which enables a better representation of the nuclear regions. Sources
are also described individually in the Appendix. The average radial
profile is also shown in Figure~1, with either the best-fit
decomposition into PSF plus host galaxy, or the PSF alone for
unresolved sources. A total of \nres sources were resolved out to
$z\sim0.6$. In particular, with few exceptions, all \low_z sources at
$z<0.5$ were resolved. Both 0851+202 ($z=0.306$) and 0954+658
($z=0.367$) have an extremely bright central point source and it is
not surprising that the host remains undetected in our short
images. The latter also has uncertain redshift and may actually be
more distant. In principle, there may be one more unresolved source
at $z<0.5$, 0735+178, for which only a lower limit to the distance is
measured ($z>0.424$).

For $z>0.5$ our success rate is much lower, with only \res_high_z of
\high_z sources resolved. This is not due to a lack of spatial
resolution but to the fact that for $z>0.6$, the F702W filter is
mapping the host galaxy spectrum short-ward of the 4000\AA\ break, so
snapshot exposure times are no longer sufficient to detect the rapidly
dimming host. Moreover, as the redshift increases, the nuclear
component becomes brighter because our sources are selected from
flux-limited samples, so contrast with the host increases unless the
host galaxies are unusually luminous.

Finally, three resolved objects have no nuclear point source 
(0145+138, 0446+449 and 0525+713). Because no nuclear activity
was found, their identification as BL Lac objects is dubious
(see notes in the Appendix) and we focus on the remaining
\resolved host galaxies.

To help understand these results, we show in Figure~4 the apparent
magnitudes of the nucleus versus the host galaxy (computed assuming
elliptical morphology if the host is unclassified). Resolved BL Lac
objects fill a 4~mag-wide band inclined at $\sim45$~degrees,
unresolved objects cluster in the bright-nucleus/faint-galaxy area,
and there are no BL Lacs in the bright-galaxy/faint-nucleus zone
(where they could easily be resolved). Clearly, the resolved BL Lac
objects lie in this central band because of our ability to resolve
only those host galaxies with luminosity within few magnitudes of the
point source. Fainter galaxies generate the observed upper limits.

Unresolved objects do not follow the same trend. These are all
distant sources ($z\gtrsim0.5$), included in the original flux-limited
samples because of their intrinsically bright nuclei (see Figure~2 in 
Paper II). Indeed, the BL
Lacs studied here come either from the optically-selected PG sample or
from X-ray- and radio-selected samples with an effective optical flux
limit of $m_V \lesssim 20$~mag due to spectroscopic identification.
In contrast, the host galaxy had little or no effect on original
source selection. Thus the systematically brighter nuclei in the high
redshift sources are more likely to out-shine their host galaxies.

The lack of BL Lac objects in the region where the host is
most easily detected is easily explained as a consequence 
of AGN classification. Radio or X-ray sources with a bright nucleus 
are classified as BL Lac objects, while those with a weak 
nucleus and prominent host are classified as galaxies. Specifically, for a 
radio source to be classified as BL Lac, the contrast of the 
4000\AA\ break must be smaller than 25\% (Dressler \& Shectman 1987;
Stocke \ea 1991; Owen \ea 1996). Assuming an intrinsic break of 
50\% for an elliptical galaxy, and a power-law spectrum for the 
non-thermal component with spectral index $\alpha=-1.5$
($F_\nu \propto \nu^\alpha$), close to the steepest values 
reported for BL Lacs (Falomo, Scarpa \& Bersanelli 1994;
the limit would be more severe for flatter power laws),
sources with $m_{host}<m_{nucleus}+1.3$ are classified as galaxies.
Indeed, radio galaxies fill the lower right-hand
corner of the diagram in Figure~4.
A few BL Lac objects stray into the galaxy ``zone,'' likely because of
nuclear variability. Although it is obvious, it is worth noting that
the average properties of separate AGN classes will be different even
when the separation is arbitrary and the properties are continuous
across the boundary (e.g., Scarpa \& Falomo 1997). In fact, we see a
continuous range of absolute nuclear luminosities going from radio
galaxies to BL Lac objects (Paper II).

\subsection{Comparison With Previous Results}

In recent years several authors have published magnitudes of BL Lac
host galaxies. The Canada-France-Hawaii (Wurtz, Stocke \& Yee 1996),
and ESO-NTT (Falomo 1996) imaging surveys of BL Lacs represent the two
largest data sets obtained from the ground. There are 24 and 8 objects
in common with ours, respectively, so a detailed comparison is possible.

For the host galaxy magnitudes we found good agreement, with modest
differences (Figure~5; Wurtz, Stocke \& Yee magnitudes were
transformed from Gunn r to Cousins R band assuming $r-R=0.3$~mag).
The average magnitude difference is $<m_{HST}-m_{ground}>=0.1$~mag,
well within our estimated systematic uncertainty, in spite of all the
transformations from the different photometric systems involved. The
dispersion is $0.4$~mag, showing that the object-to-object discrepancy
can be substantial (see the Appendix for one-to-one comparisons).

For the effective radius, the agreement is not as good, probably
because this parameter is not as precisely determined in the fit.
This is because $r_e$ depends strongly on the radial profile at large
radii, where the signal-to-noise ratio is lowest.
The measured values of $r_e$ do not correlate with 
the intensity of the 
background, meaning at least statistically our determination 
of the background is correct. The total
magnitude of the host is, however, only marginally affected by a 
wrong estimate
of $r_e$, because only a minor fraction of the total light is
contained in the external regions of the galaxy. For instance, for the
BL Lac object 0414+009, observed in both the CFHT and NTT surveys, the
estimated effective radii are very different while the total magnitude
of the host galaxy is basically the same (and also equal to our
value). The average difference between published measures of $r_e$ and
ours is 0.2~kpc, only $\sim2.5$\% of our median $r_e$, sufficiently
small that we conclude there are no systematic deviations. The dispersion
is much larger, 9~kpc, clearly showing the large uncertainty in
determining this parameter.

\subsection{Host Galaxy Morphology}

BL Lac objects are hosted almost universally by elliptical galaxies
(Ulrich 1989; Wurtz, Stocke \& Yee 1996; Falomo 1996; Kotilainen,
Falomo \& Scarpa 1998). However, in a small number of cases
the host galaxy has been classified as a disk system (Halpern \ea 1986;
Abraham, McHardy \& Crawford 1991; Wurtz, Stocke \& Yee 1996). 
It has been argued that these detections of S0 or spiral host galaxies 
invalidate the unified model for BL Lacs (see discussion in
Urry \& Padovani 1995). Thanks to its superior spatial
resolution, $HST$ is at present the best instrument to distinguish between
elliptical and spiral hosts. This is because the radial profiles of the two
types of galaxies differ mainly in the center, where ellipticals are
substantially more peaked than disks. 
Ground-based observations usually resolve the host galaxy only at
larger radii, where the two galaxy models have similar slopes.

To show how clearly $HST$ data can discriminate a \dvac from a disk
model we show two representative examples in Figure~6. On the left is
0607+711, which was thought to be hosted by a disk galaxy (Wurtz,
Stocke \& Yee 1996; ground-based study of this source is hampered by a
bright star in the field of view). With $HST$ we can trace the host
galaxy surface brightness profile very close to the nucleus, where the
exponential law is completely inadequate to describe the observed
profile. Thus we rule out the disk model with high confidence, even
though the external part of the radial profile is quite noisy because
of the nearby star.

The other example is 1418+546 (OQ~530), which
at least three different groups reported was in a disk galaxy
(Abraham, McHardy \& Crawford 1991; Stickel
\ea 1993; Wurtz, Stocke \& Yee 1996). This time the discrimination
between the two models is not easy even at $HST$ resolution.
The exponential law indeed gives a reasonably good fit, but fails to
describe the inner 1.5~arcsec of the observed radial profile (hence it
is not surprising that from the ground it was classified as a disk
galaxy). In our data, the \dvac model is preferred at the 99\%
confidence level. For a composite disk-plus-bulge model, bulge
and disk have similar luminosity. Given
the higher mass-to-light ratio of the bulge, the system would
still be bulge dominated.

These two cases illustrate the results characteristic of all
well-or moderately well-resolved objects. 
The disk model is always ruled out because it fails to
describe the innermost part of the radial profile.

For the sample as a whole, we found unambiguous morphological
classifications for \cls hosts, while \ucls more hosts could not be
classified definitely in one class or the other. For these sources
deeper observations are obviously required. All classified hosts but
one are ellipticals. The only object that resides in a disk host is
0446+449, which has no nuclear point source and therefore is a galaxy
rather than a BL Lac object. Another BL Lac object that appears to be
in a disk galaxy is PKS~1413+135 (Abraham, McHardy \& Crawford 1991;
Wurtz, Stocke \& Yee 1996), which was not in our $HST$ sample. Also
in this case there is no sign of a nuclear point source, and it has
been suggested that the BL Lac lies behind a foreground disk galaxy,
or that it is not a BL Lac at all (Stocke \ea 1992; Perlman \ea 1994;
McHardy \ea 1994, 1991), Given the lack of nuclear activity in
0446+449 and 1413+135, their identification as BL Lacs is highly
questionable.

Three other disk hosts have been reported but none has held up. These
include 1415+259 (Halpern \ea 1986); 1418+546 (Abraham, McHardy \&
Crawford 1991; Wurtz, Stocke \& Yee 1996), 
which with 1413+135 implied (at the time) that
the fraction of disk galaxies was large; and MS~0205+351 
(Wurtz, Stocke \& Yee 1996).
However, 1415+259 was later reclassified as an elliptical (Romanishin
1992; Wurtz, Stocke \& Yee 1996), 
the host galaxy of 1418+546 has been shown here to be
elliptical, PKS~1413+135 is most probably not a BL Lac object, and
MS~0205+351 was reclassified as an elliptical (Stocke, Wurtz \&
Perlman 1995; Falomo \ea 1997). Hence, at present there is not a
single, unquestionable detection of a disk host galaxy.

In our large $HST$ data set, there is not a single BL Lac object
hosted by a disk-dominated galaxy. Combining our sample with the CFHT
survey, a total of 66 BL Lac hosts have been classified as
ellipticals. The probability of randomly extracting a sample of 66
elliptical host galaxies from a mixed distribution is quite low. At
99\% confidence, the fraction of disk systems must be $<7$\% of the
total population. That so few host galaxies can be spirals, and all
host galaxies could be ellipticals, strongly reinforces the proposed
unification of BL Lac objects and radio galaxies.

\subsection{Host Galaxy Apparent Magnitudes and Parent Population}

The apparent magnitudes of the BL Lac host galaxies increase with
redshift as expected for fixed luminosity, as shown in the Hubble
diagram in Figure~7. 90\% of the values are within 1~mag of the line
corresponding to the median absolute magnitude, $M_R = -23.7$~mag
(for $H_0=50$ km/s/Mpc and $q_0=0$).
This holds up to redshift $\sim 0.6$, beyond which there are mainly
lower limits to the apparent magnitude. This suggests that BL Lac
host galaxies are quite similar in luminosity, as they are in
morphology. A simple passive evolution model for elliptical galaxies
formed at high redshift (Bressan \ea 1994) is roughly consistent with
the observations; the predicted evolution is in any case very small at
$z\lesssim 0.6$. The mean luminosity of the detected host galaxies
does increase slightly with redshift, but this is expected at least
qualitatively because of several selection effects. More data at
$z\gtrsim0.6$ are necessary to properly address the evolution of the
host galaxies.

We now compare the BL Lac host galaxies to three samples of radio galaxies.
The first includes FR~I and II radio galaxies 
from a subset of the 2~Jy radio sample (Morganti \ea 1993), 
with redshifts and magnitudes as reported by Wall \& Peacock (1985). 
This sample is interesting because 
it was selected with a similar radio flux limit as 
the 1~Jy sample of BL Lacs (Stickel et al. 1991), 
radio morphologies of the sources are well defined, and
it covers essentially the same redshift range as the BL Lac objects.
The top plot in Figure~8 shows that the BL Lacs host galaxies
are essentially indistinguishable from the radio galaxies, 
across the full redshift range, although at any given redshift
the dispersion of radio galaxy magnitudes is larger 
(because of selection effects explained in \S~3.1).

The other two samples of radio galaxies
include more objects but at much lower redshift.
The Ledlow \& Owen (1995) sample includes only sources 
in clusters, mostly FR~Is in the redshift range $0<z<0.25$. 
The Govoni \ea (1999) sample includes FR~Is and IIs at 
somewhat lower redshift $z\lesssim 0.12$, 
both in clusters and in the field.
The bottom plot in Figure~8 shows these radio galaxies in the
Hubble diagram. The radio galaxies again have very similar apparent
magnitudes to the BL Lac host galaxies, with the same 
larger dispersion relative to the mean at any given redshift.

This comparison is convincing evidence in favor of the unification 
of radio galaxies and BL Lacs. For more discussion, see Paper II.

\subsection{The BL Lac Nuclei}

The Hubble diagram for the BL Lac nuclei, in Figure~9,
shows much larger spread than for the host galaxies. 
The nuclear apparent magnitudes are mostly in the range 
$15\lesssim m_R \lesssim 20$, with little dependence on redshift,
as expected in flux-limited samples. (The samples plotted here have an
effective flux limit of $m_R\sim20$~mag due to the criterion for
optical identification.)
The $L_{nuc}/L_{host}$ luminosity ratio should therefore increase with $z$.
For resolved sources, $L_{nuc}/L_{host}$ varies from 0.03 to 5, 
with a median near 1, reflecting the distinction between BL Lacs and
radio galaxies (weak nuclei with bright hosts) and/or
our inability to resolve hosts much fainter than the nucleus
(see \S~3.1).

In \S~3.4 we showed that the host galaxies have nearly constant
luminosity with redshift (within $\pm 1$~mag). At high redshift,
where many BL Lacs are unresolved, we estimate $L_{nuc}/L_{host}$ by
varying the (assumed) host galaxy magnitude by $\pm 1$~mag from the
sample average. The ranges of expected $L_{nuc}/L_{host}$ are shown
in Figure~10 as a function of redshift, along with the measured values
for resolved sources. This Figure suggests that $L_{nuc}/L_{host}$
increases significantly with distance, which reflects the bias induced
by the flux limit of the BL Lac sample. This means it is misleading to 
compare this quantity among samples covering different redshift ranges.

\section{Conclusions}

We reported on a large $HST$ survey of BL Lac objects, focusing on 
the careful procedure for detection and study of the host galaxy.
Simulations show that these procedures recover the 
true source parameters to within the reported errors.
We found a simple empirical relation to represent 
the wings of the $HST$ point spread function, at 
$\gtrsim2$~arcsec from the center; properly accounting for the
scattered light at large radii is essential to not overestimating
the galaxy luminosity, and indeed to avoiding false detections.

In the full sample of \ntot objects, 
we detect \nres host galaxies, including essentially all sources 
at $z\lesssim 0.5$. 
The data set is large enough and homogeneous enough to allow for
a careful assessment of important selection effects.
We do not detect bright nuclei with very faint host galaxies,
where the PSF swamps the galaxy light, nor
do we detect faint nuclei with bright host galaxies, 
which would have been classified as radio galaxies 
rather than BL Lac objects.

Three objects, 1ES~0145+138, 1ES~0446+449 and
1ES~0525+713, are very unusual in that they lack a nuclear point source. 
The morphological types of these three galaxies were determined:
0446+449 is in a disk galaxy, and the other two are in ellipticals. 
Possibly they are BL Lac objects in a quiescent phase,
but a careful re-identification of the optical counterpart is
clearly needed.

Of the remaining \resolved host galaxies resolved, \cls were classified
morphologically, invariably as elliptical. In the
whole sample the only BL Lac residing in a disk galaxy is 0446+449,
a possible misidentification. 
We also reclassify as elliptical the host of 1418+546, previously
reported to be a disk.
We conclude that in our survey, carried out at a spatial
resolution $\sim 10$ times greater than previously possible from the
ground, there are no cases of an unquestionable BL Lac 
in a disk galaxy. This is particularly significant given
the large number of galaxies studied, the homogeneity of the data set, 
and the careful reduction procedures.
At the 99\% confidence level, this implies that
the fraction of disk galaxies must be smaller than 8\% of the 
total BL Lac populations; adding a few more host galaxies resolved and 
classified from the ground reduces this fraction to 7\%.
Our results are thus fully consistent with the hypothesis that
BL Lacs reside exclusively in elliptical host galaxies.

Comparison with previous imaging studies of BL Lac objects shows
good agreement for the total magnitudes of the host galaxies, 
as well as good agreement on average for the effective radii, although
for individual objects the disagreement can be quite large.

The apparent magnitudes of the host galaxies as a function of distance are
consistent with a roughly constant luminosity, 
with spread of $\pm 1$~mag. There are too few host galaxy detections
at $z>0.6$ to constrain possible luminosity evolution.
Finally, comparison of the Hubble diagram for BL Lac hosts and radio 
galaxies fully supports their unification.

\acknowledgements
Support for this work was provided by NASA through grant number
GO6363.01-95A from the Space Telescope Science Institute, which is
operated by AURA, Inc., under NASA contract NAS~5-26555, and by the
Italian Ministry for University and Research (MURST) under grant Cofin98-02-32. 
We thank Jessica Kim for her research on the BL Lac redshifts.
RF thanks the $HST$
visitor program for hospitality during several visits to STScI. 
This research made
use of the NASA/IPAC Extragalactic Database (NED), operated by the Jet
Propulsion Laboratory, Caltech, under contract with NASA, and of
NASA's Astrophysics Data System Abstract Service (ADS).

\appendix{Results on Individual Objects}

\noindent{\bf 0033+595}: Two objects, separated by $1\sec .58$, are found 
close to the possible optical counterpart of the BL~Lac. From 
the $HST$ image alone we could not decide which one is actually the AGN, but 
based on source color and radio emission the southernmost of the two 
is the most probable optical counterpart of 
this BL Lac object (Scarpa \ea 1999a). Both objects are unresolved with $HST$.\\

\noindent{\bf 0118-272}: The nuclear source of this distant BL Lac 
object ($z>0.559$)
is very bright ($M_R=-27.4$~mag). Not surprisingly, the host galaxy
is not detected. Some small galaxies are visible
in the PC field of view, including one $1\sec .56$ from the nucleus,
at $PA=71^\circ$.\\

\noindent{\bf 0122+090}: The host galaxy is fully resolved. The radial profile is
well described by a \dvac law, plus a modest contribution
from a nuclear point source. 
For the host we measure $m_R=18.9$~mag and $r_e=1\sec .05$,
similar to the values found by \wsy, $m_R=18.35$~mag and $r_e=1\sec .6$.\\

\noindent{\bf 0138-097}: The radial profile is largely compatible with the PSF,
but a clear excess ($>1$~mag) is visible at large radii.
This excess of light may be due either to the underlying host galaxy, or
to the nearby galaxies that surround the BL Lac. Increasing the sky by 1$\sigma$,
the radial profile is consistent with the PSF, and we conclude the 
object remains unresolved. A companion galaxy is visible 1.44~arcsec
from the BL Lac and 3 more fainter galaxies are detected within 5~arcsec.\\

\noindent{\bf 0145+138}: As is the case for two other objects from the Einstein
Slew survey (0446+449 and 0525+713), no indication of a nuclear point
source is found in this BL Lac. The offset of the Slew X-ray source
from the proposed optical counterpart is quite large ($\sim 150$~arcsec) 
and the source was not detected in the Rosat all-sky survey. 
The absence of nuclear activity, along with the
lack of any indication of a power-law continuum in the optical band
(see the spectrum in Perlman \ea 1996), makes the identification
suspicious. The radial profile of the galaxy follows perfectly 
a de~Vaucouleurs law.\\

\noindent{\bf 0158+003}: Clearly resolved with a radial
profile very well described by an elliptical host galaxy plus a nuclear
point source. A large galaxy is
visible $2\sec .3$ north-west of the BL Lac. It is characterized by a
large dust lane, and is most probably an edge-on spiral.\\

\noindent{\bf 0229+200}: This relatively close source is fully resolved in our
$HST$ image. The host galaxy is elliptical with a radial profile
well described by a \dvac law. A disk model is ruled out.
No companion are visible and the host appears to be absolutely normal.\\

\noindent{\bf 0235+164}: This distant object is unresolved. Our image is not deep enough
to enable a full discussion of the very interesting environment that 
characterizes this BL Lac object (e.g., Falomo 1996). 
We clearly detect only the southern source located 
1.99~arcsec from the BL Lac at position angle $175^\circ$.
It is unresolved and, given the reported redshift $z=0.524$, is most
probably an AGN.\\

\noindent{\bf 0257+342}: The host galaxy is elliptical with total magnitude 
$m_R=17.9$~mag and $r_e=1\sec .75$. Our values agree very well
with the measurement by \wsy, $m_R=17.9$~mag and $r_e=2\sec .0$.\\

\noindent{\bf 0317+178}: The object is characterized by a bright nucleus surrounded
by a round elliptical host, and a companion galaxy $2\sec$ south.
The host radial profile follows well a \dvac law, with best-fit parameters 
$m_R=17.6$~mag and $r_e = 3\sec .25$, moderately larger than 
found by \wsy, $m_R=18.0$~mag and $r_e = 1\sec .2$.
A disk model for the host is ruled out.\\

\noindent{\bf 0331-365}: The source is well resolved into a point source
surrounded by a round elliptical host. A disk model is ruled out.\\

\noindent{\bf 0347-121}: The source is clearly resolved with radial profile
well described by a point source plus a \dvac law. 
A disk model for the host galaxy is not acceptable.
The optical image shows a large
interacting system of three galaxies 11.7~arcsec north of the BL Lac. 
However, there are no indication of connection with the BL Lac.\\

\noindent{\bf 0350-371}: At $HST$ resolution the source appears fully resolved, 
showing a rather elliptical host, which has a radial profile 
well described by a \dvac law. A disk galaxy model for
the host is ruled out. A large galaxy is located $6\sec .2$ west
of the BL Lac.\\

\noindent{\bf 0414+009}: At least four independent determinations of the host galaxy
magnitude are available for this moderately distant BL Lac object ($z=0.287$).
The two most recent measurements both find
$m_R= 17.4$~mag and $r_e \sim 5\sec$
(Falomo 1996; Wurtz Stocke \& Yee 1996).
The $HST$ image clearly shows the host galaxy surrounding a very bright
point source, with a radial profile well described by a
point source plus an elliptical galaxy. 
Our best-fit parameters for the host are $m_R=17.5$~mag and $r_e=4\sec .7$,
in good agreement with previous measurements. Despite being the
dominant member of a moderately rich cluster (Abell class 0; McHardy
\ea 1992; Falomo, Pesce \& Treves 1993), 0414+009 appears 
isolated in our image.\\

\noindent{\bf 0419+194}: This distant ($z=0.512$) source is resolved.
The radial profile is better fitted by a \dvac law, with best-fit
parameters in the usual range for BL Lac hosts and confirming the
values reported by Wurtz, Stocke \& Yee (1996).\\

\noindent{\bf 0426-380}: This distant object ($z>1.03$) remains unresolved.
A very small companion galaxy is visible 
$0''.51$ from the BL Lac, at P.A.=$260^\circ$.\\ 

\noindent{\bf 0446+449}: The proposed optical counterpart of this X-ray-selected 
BL Lac object is a normal disk galaxy with no nuclear point source. 
This is similar to what is found for two
other objects (0145+138 and 0525+713) from the Einstein Slew survey.
The coordinates of the proposed optical
counterpart (Table~1) differ by $\sim 10\sec$ from the VLA
radio coordinates $\alpha = 04:50:07.2 , \delta = 45:03:12$ (Perlman
\ea 1996). Interestingly, none of the several point sources close to the
proposed optical counterpart match the VLA coordinates. The
original coordinates of the X-ray source are +191,+114~arcsec apart
from the VLA position (and therefore also from the optical ones) making the 
identification highly uncertain. \\

\noindent{\bf 0454+844}: The redshift of this source has recently been 
determined by Stocke \& Rector (1997) to be $z>1.340$, completely 
different from the earlier value ($z=0.112$;
reported by Stickel \ea 1994 as a private 
communication by C. Lawrence). 
Our observation is fully consistent with the object being at high
redshift, with no indication of the host galaxy and
a radial profile fully consistent with our PSF model.\\

\noindent{\bf 0502+675}: This source turns out to be double: two objects with
comparable magnitude and separation of only 0.33~arcsec are clearly
visible at $HST$ resolution. This is a promising gravitational
lens candidate (Scarpa \ea 1999a). The brightest 
source is resolved; however, we were unable to discriminate 
between different host models.\\

\noindent{\bf 0506-039}: Unfortunately, the BL Lac
object falls partly on the PC and partly on the WF camera.
We were still able to extract a reliable radial profile,
finding the source is fully resolved and the host well described
by a de~Vaucouleurs law.\\

\noindent{\bf 0521-365}: This well studied nearby object is fully resolved. The radial 
profile extends up to 14~arcsec from the center and 
is perfectly described by a \dvac law plus a point source. 
The optical jet is also clearly visible and can be studied as close as 
0.3~arcsec from the nucleus (Scarpa \ea 1999b).\\ 

\noindent{\bf 0525+713}: This relatively nearby object ($z=0.249$) is fully
resolved. Unfortunately, due to incorrect coordinates, the object
falls on the edge of the WF2 and extends partially on WF3, rather than being
in the center of the PC camera. Its radial profile
looks like that of a normal elliptical galaxy, well described by a
simple \dvac law. As is the case for other two objects from the
Einstein Slew survey (0145+138 and 0446+449), no indication of nuclear activity
was found. The coordinates of the proposed optical counterpart (Table~1), as
identified from the published finding chart, differ by $\sim28''$ 
from the VLA coordinates (Perlman \ea 1996), much more 
than the stated precision of the VLA radio
coordinates ($\sim 1''$). Therefore either there is a mistake in
the reported Slew coordinates or in marking the optical counterpart in
the finding chart. The absence of a point source at the center of the
galaxy together with the lack of any indication of a power-law
continuum in the optical band (see the spectrum in Perlman \ea 1996),
makes the identification suspicious.\\

\noindent{\bf 0537-441}: This distant ($z=0.896$), bright BL Lac is not
resolved. Our 10-minute exposure is largely saturated in the
center, allowing the study of the PSF wing up to 7~arcsec from the 
center. From Figure~1 one can get an idea of the amount of scattered
light present in the PC camera (see also the cases of 1553+113,
1424+240). The comparison of this profile with our PSF model shows a
perfect match, confirming the object is unresolved and
our PSF model correct. Some companion galaxies are visible.\\

\noindent{\bf 0548-322}: This well-studied BL Lac is fully resolved. The host is
a large elliptical with radial profile well described by a \dvac law.
Our measurements for the host galaxy, $m_R=14.62$~mag and $r_e=7\sec .05$,
agree exactly with what was found by \wsy, 
$m_R=14.6$~mag and $r_e=7\sec .7$, but are quite different from the values
reported by Falomo, Pesce \& Treves (1995), 
$m_R=14.1$~mag and $r_e=28\sec$. The reason for 
such a discrepancy is not clear.\\

\noindent{\bf 0607+710}: This relatively nearby object is clearly resolved.
Despite the presence of a bright star that makes
the extraction of the radial profile difficult, we can follow it 
out to 6~arcsec from the center. It is 
described by a nuclear point source plus an
elliptical host galaxy. Contrary to what was found by \wsy,
who reported the host to be a spiral, we are able to rule
out a disk model (Fig.~5). 
Their result may have been influenced by the presence of the bright star.\\

\noindent{\bf 0622-525}: There is no redshift information for this clearly resolved 
source. The radial profile is well
described by an elliptical host galaxy plus a nuclear point source,
while a disk model is ruled out.\\

\noindent{\bf 0647+250}: Our 10-minute $HST$ exposure of this BL Lac shows a
bright point-like object. The image is well exposed and the 
radial profile can be traced out to 6~arcsec, to a faint surface brightness,
$\mu_R \sim 26$~mag~arcsec$^{-2}$. 
The agreement with the PSF model is very good, enabling
us to put an interesting lower limit to the host apparent magnitude 
$m_R>19.1$~mag. Including 0.35~mag of K correction, and conservatively 
assuming the host has $M_R=-22.7$~mag, i.e., 1~mag fainter 
than the median for this sample, our upper limit corresponds
to a minimum distance for the source of $z\gtrsim 0.3$.\\

\noindent{\bf 0706+591}: This nearby source ($z=0.125$) is fully resolved. The
radial profile is well described by an elliptical profile plus a
nuclear point source, a disk model being definitely ruled out.
Within the halo of the BL Lac host galaxy, 4.2~arcsec from the nucleus
(projected distance of 12.8~kpc), is a companion disk galaxy
with integrated total magnitude of $m_R= 21$~mag and r$_e=0\sec .3$. 
Some other small galaxies, probably background objects, are
present in the PC field of view.\\

\noindent{\bf 0715-259}: The object is resolved into a point source surrounded 
by a small, rather elongated host galaxy. We were
not able to discriminate between disk or \dvac models.\\

\noindent{\bf 0716+714}: This BL Lac at unknown distance is a 
very bright object for which the radial profile can be 
studied over a large range of surface brightness. 
The agreement with our PSF model is excellent.\\

\noindent{\bf 0735+178}: For this source only a lower limit to the distance is 
available ($z>0.424$; Carswell \ea 1974).
The apparent magnitude of the nucleus (V$\sim 16.5$~mag)
implies it is particularly luminous, which may
explain why the host galaxy is not detected. If
0735+178 were at $z=0.424$, our lower limit for the host galaxy apparent magnitude
requires an absolute K-corrected magnitude $M_R >-22.7$~mag.\\

\noindent{\bf 0737+744}: The object located at $z=0.315$ is resolved. The radial 
profile is described by a point source plus an elliptical host galaxy.
A disk model is ruled out.\\ 

\noindent{\bf 0749+540}: There is no published redshift for this source,
unresolved in our $HST$ image. This suggests it may actually be quite 
distant. To be conservative, assuming the host has 
$M_R=-22.8$~mag, i.e., 1~mag fainter than the average measured 
for this sample, and including a K correction of 1~mag, 
our limit $m_R>21.9$~mag corresponds
to quite a large minimum distance of $z\gtrsim 0.7$.\\

\noindent{\bf 0754+100}: This BL Lac was resolved previously from the ground
(Abraham, McHardy \& Crawford 1991; Falomo 1996). 
The estimated absolute magnitude
for the host galaxy corresponds to an extremely bright galaxy ($M_R\sim
-25$~mag). This result put in question the available redshift ($z=0.66$)
tentatively reported by Persic \& Salucci (1986). A galaxy visible
13.6~arcsec north-east of the BL Lac object is known to be at $z=0.27$
(Pesce, Falomo \& Treves 1995). 
If the host galaxy were at the same redshift, it would have a
more reasonable absolute magnitude and would be expected to be
resolved from the ground. At $HST$ resolution 
the source appears slightly resolved. The excess over the PSF model can
be tentatively described with a \dvac model, obtaining an
apparent magnitude for the host of $R\sim19.2$~mag and r$_e \sim2''$. 
These figures are in complete agreement with what was found by Falomo 
(1996) from subarcsecond ground-based imaging ($R\sim18.9$~mag, r$_e \sim2''.3$).
However, based on our statistical criteria the source remains unresolved
and for its host galaxy only an upper limit is reported in Table~3.\\

\noindent{\bf 0806+524}: The redshift for this source has been recently
measured by Bade \ea (1998), who found the source is moderately
nearby ($z=0.138$). At $HST$ resolution 0806+524 is fully resolved and
an interesting arc-like structure $1''.93$ south of
the nucleus is seen (Scarpa \ea 1999a). 
The radial profile of the host is described by a point source plus an
elliptical host. A disk model is ruled out.\\

\noindent{\bf 0814+425}: The reported redshift $z=0.258$ by Wills \& Wills (1976)
is based on two weak emission lines identified with
MgII(2798\AA) and [O II](3727\AA). 
The same authors state that a second spectrogram of apparently
similar quality failed to confirm these lines. Later, the spectrum was
found featureless on at least two occasions (Dunlop \ea 1989; Stickel \ea
1993) so that the proposed redshift has never been confirmed. More
recently Lawrence \ea (1996) obtained a very high signal-to-noise
spectrum, which shows three tiny features at discordant redshift
(either 0.245 or 1.25), neither the same as 
the previously proposed redshift. We therefore assume the
redshift is still unknown. Our $HST$ image shows a point-like
object, with radial profile well described by the adopted PSF
model. Conservatively assuming the host has $M_R=-22.8$~mag, 
i.e., 1~mag fainter than the
average measured for this sample, our upper limit on the host galaxy apparent
magnitude ($m_R>20$~mag including 0.9~mag of K correction) corresponds
to a minimum distance of $z\gtrsim 0.5$.\\

\noindent{\bf 0820+225}: The profile of this distant object ($z=0.951$) is 
stellar. No companions are visible on the frame.\\ 

\noindent{\bf 0823+03}: The presence of a bright star close to the
BL Lac object severely hampers its study. To minimize the impact of the
scattered light, we averaged the radial profile only in the 180$^o$
opposite the star. The resulting profile lies slightly above 
the PSF, suggesting the source may be resolved. However, increasing the
sky level by $1\sigma$, the source is no longer resolved; hence, we
conservatively assume it remains unresolved.
This BL Lac was possibly resolved by \wsy, who estimate an absolute
magnitude for the host galaxy of $M_R=-23.9$~mag.\\

\noindent{\bf 0828+49}: The source is resolved, the radial profile being significantly
above the PSF. We were not able to discriminate between a disk or
elliptical morphology for the host galaxy. The \dvac law
best fit corresponds to $m_R=20.26$~mag, the same as
reported by \wsy.\\

\noindent{\bf 0829+046}: This BL Lac object was resolved from the
ground by Abraham, McHardy \& Crawford (1991) who found $m_R=16.7$~mag and
r$_e=23\sec $, and by Falomo (1996), who found $m_R=17.6$~mag and
r$_e=4\sec .8$. The source is fully resolved in our $HST$ image; however,
due to a failure of the telescope tracking, the image is slightly
elongated and we were not able to fit properly the central part of
the average radial profile (within $1\sec$ from the center). To 
describe the profile center properly, we created a PSF model using a star
visible in the PC camera. The star was not very bright, permitting us
to follow the radial profile only out to $\sim2$~arcsec from the
center. Because the wings of the PSF profile are only marginally
affected by the poor tracking, we smoothly joined the star's radial profile
to the PSF model used for the other objects.
Using this customized PSF we were able to describe the observed radial
profile quite well. The best-fit \dvac law
gives $m_R=16.9$~mag and r$_e=4\sec.3$, substantially brighter than 
reported by Falomo (1996).
The two measures of the effective radius, however, agree well. 
In contrast, Abraham, McHardy \& Crawford (1991) report $r_e=22\sec$, 
a value well outside our estimated errors and also much larger than
found for any other elliptical galaxy; the fit also seems poor in
their Figure~1. 
We could not discriminate between disk and \dvac models.\\

\noindent{\bf 0851+202}: This well-known BL Lac objects was
imaged with $HST$ by Yanny, Jannuzi \& Impey (1997), who reported it was 
resolved. However, they identify the host with a diffuse structure 
$\sim 0.4$~arcsec off-center with respect to the BL Lac nucleus. The same structure was
not detected in a much deeper image obtained with the Nordic Optical Telescope
(Sillanp\"a\"a \ea 1998), and we suggest it is a ghost image produced within
the WFPC2 camera by the
bright nucleus. The host was possibly detected by \wsy, 
who report $m_R=19.8$~mag. The average radial
profile derived from our $HST$ image, which was a short exposure
of only 5 minutes, shows a perfect match with the PSF model.
This is the nearest source ($z=0.306$) that remains unresolved.\\

\noindent{\bf 0922+74}: This moderately distant source ($z=0.638$) is resolved.
The host is an intrinsically very bright elliptical ($M_R=-24.6$~mag
including K correction of 1.34~mag). A disk model is ruled out. From 
the ground, the host galaxy has never been clearly detected.
\wsy\ report a marginal detection with
$m_R\sim 21.9$~mag, more than a magnitude fainter than our
measurement $m_R\sim 20.25$~mag.\\

\noindent{\bf 0927+500}: The BL Lac object is fully resolved into a point source 
surrounded by a round elliptical host. A disk model is ruled out. Several small
galaxies are visible in the frame, suggesting the source may be member
of a small cluster.\\

\noindent{\bf 0954+685}: This source is among the nearest ($z=0.367$) in our sample that
remain unresolved. The lower limit to the host galaxy apparent magnitude
is $m_R\sim 19.6$~mag, a value that coincides with that
found by \wsy, who claimed it was marginally resolved. 
The fact that this moderately nearby
source was unresolved with $HST$ is most probably due to 
the brightness of the nucleus ($m_R\sim 16.1$~mag) during 
our observation.
We note, however, that the redshift deserves some comment.
There are two determinations, both quite uncertain. 
Lawrence \ea (1986) report the marginal detection of very faint
[OIII]5007\AA\ emission and Mg~b band absorption lines at $z=0.368$.
However, they concluded that the redshift is a ``tentative value,
requiring confirmation.'' 
Later, Stickel, Fried \& K\"uhr (1993) detected
[OII] 3727\AA\ and CaII 3933,3968\AA\ at the same redshift ($z=0.367$), 
but no one line has been detected twice. It would therefore be
highly desirable to have a firmer determination of the distance to
this object.\\

\noindent{\bf 0958+21}: The BL Lac object is
resolved and surrounded by several galaxies. Among 
them is a big nice spiral. The nuclear point source
was very faint during the observation, allowing a detailed 
study of the host galaxy. The radial profile is 
consistent with a simple \dvac law, while a disk model is ruled out.\\

\noindent{\bf 1011+496}: The redshift of this object is uncertain, being
based on the possible membership of the BL Lac in the cluster Abell~950, 
and some galaxies are indeed detected in the field of view,
at $z=0.20$ (Wisniewski \ea 1986). 
In our $HST$ image the source is clearly resolved, with 
an average radial profile well described by a point source plus an
elliptical host galaxy. A disk model for the host is not acceptable.\\

\noindent{\bf 1028+511}: A reliable redshift of $z=0.361$ based on CaII H\&K
absorption lines has been recently reported by Polomski \ea (1997), 
a value somewhat larger than the previously reported $z=0.239$.
The object is resolved, and its radial profile is well
described by a \dvac law plus a nuclear point source.\\ 

\noindent{\bf 1044+54}: The BL Lac object is resolved and the radial profile is well
described by a nuclear point source plus a \dvac
law. There is no reported redshift for this
object. Conservatively assuming the host has 
$M_R=-24.8$~mag, i.e., 1~mag brighter than the average measured 
for this sample, and including a K correction of 1.4~mag, 
the measured apparent magnitude $m_R=20.05$~mag implies $z<0.65$.\\

\noindent{\bf 1104+384}: The source is fully resolved with $HST$. The host galaxy is
elliptical with quite normal isophotes, indicating no
tidal stress due to the large companion galaxy $14''.2$ to the north-east. The
nuclear point source is perfectly centered on the host galaxy.
A disk model is ruled out. \\

\noindent{\bf 1106+24}: The BL Lac object is resolved with
radial profile well described by a point source 
plus a \dvac law; however, a disk model for the host cannot be
ruled out. The redshift of the object is unknown. We set a tentative upper 
limit for the distance assuming, to be conservative, the host has luminosity
$M_R=-24.8$~mag, i.e., 1~mag brighter than the average measured 
for this sample, and including a K correction of 1.4~mag, 
the measured apparent magnitude $m_R=19.6$~mag implies $z<0.6$.\\

\noindent{\bf 1133+168}: In our $HST$ image the BL Lac object
is very close to the PC edge, 
preventing a complete study of its radial profile. 
Despite this problem and the
moderately large redshift ($z=0.460$, Fichtel \ea 1994), the source 
appears fully resolved and well described by a point source plus 
an elliptical host galaxy. A disk model is ruled out.
Several other galaxies are visible on the WF camera.\\ 

\noindent{\bf 1136+704}: This well-studied nearby source is fully resolved.
The radial profile is perfectly described 
by a \dvac law plus a nuclear point source. The 
magnitude of the point source is ill defined as our profile is 
largely saturated on the center.\\ 

\noindent{\bf 1144-379}: This distant source ($z=1.048$) is unresolved.\\ 

\noindent{\bf 1147+245}: The BL Lac object is unresolved, its radial profile
being fully consistent with the PSF model. No
redshift information is available. To be conservative, assuming the host 
galaxy has $M_R=-22.8$~mag, i.e., 1~mag fainter than the average measured 
for this sample, and including a K correction of 0.9~mag,
our limit $m_R>20.7$~mag for the host apparent magnitude 
corresponds to a tentative minimum distance of $z>0.45$.\\

\noindent{\bf 1207+394}: At $HST$ resolution this moderately distant source is 
resolved. The host is elliptical with best-fit parameters $m_R=20.30$~mag
and $r_e=1\sec .2$, corresponding to a very bright host $M_R=-24.4$~mag. 
A disk model is ruled out.\\

\noindent{\bf 1212+078}: This nearby source ($z=0.136$) is fully resolved.
The host is a large, round elliptical. A disk model 
is definitely ruled out.\\ 

\noindent{\bf 1215+303}: The host galaxy is a large, round elliptical. A disk model 
is ruled out. Two galaxies are detected superimposed on
the host galaxy, at projected distance of $1\sec .43$ and $2\sec .76$,
together with several other small galaxies located all around
the BL Lac object. \\

\noindent{\bf 1218+30}: The BL Lac object is well resolved and the host 
galaxy is an elliptical. 
The clear detection of the host galaxy is consistent with the source 
being at the recently determined redshift $z=0.182$ (Bade \ea 1998). 
No companion galaxies were detected.\\ 

\noindent{\bf 1221+245}: This source was observed in the I band with $HST$ by 
Jannuzi, Yanni \& Impey (1997) with exposure time longer than 
the F606W (V band) snapshot observation presented here. The source is 
resolved in both cases and the derived parameters for the host galaxy agree well
with the I-band values (see Urry et al. 1999). 
The color $V-I=1.8$~mag. is fully consistent with the expectation (1.6~mag) 
for a normal elliptical galaxy at $z=0.218$.\\

\noindent{\bf 1229+643}: This nearby BL Lac object ($z=0.164$) is fully
resolved. The elliptical host looks quite symmetric despite the presence
of a companion galaxy located $3\sec .36$ to the south-west.
A disk model for the host is ruled out.
Our best-fit \dvac model corresponds to $m_R=16.4$~mag and $r_e=2\sec .0$,
which agrees well with what was found by \wsy, $m_R=16.5$~mag
and $r_e=2\sec .9$\\

\noindent{\bf 1239+069}: This BL Lac from the Einstein Slew survey remains
unresolved. We note that in our $HST$ survey we resolve 
all sources at $z\lesssim 0.3$, and the detected host galaxies show only a
small dispersion in luminosity. It is therefore surprising that this source, 
reported at $z=0.150$ (Perlman \ea 1996), remains unresolved.
At that distance, our upper limit for the host luminosity of $m_R>22.3$~mag would
correspond to an absolute magnitude $M_R=-17.6$~mag, 
extremely weak for a BL Lac host galaxy.
Because the available redshift is based on the detection of very weak lines,
it may be erroneous. In absence of a more reliable determination of the 
distance, we have assumed it is unknown.\\

\noindent{\bf 1246+586}: This BL Lac at unknown distance is unresolved. 
Conservatively assuming the host has $M_R=-22.8$~mag, 
i.e., 1~mag fainter than the
average measured for this sample, our lower limit on the host apparent
magnitude ($m_R>20.2$~mag including 1~mag of K correction) corresponds
to a tentative minimum distance of $z\gtrsim 0.5$.\\

\noindent{\bf 1248-293}: The source is clearly resolved. The host is elliptical,
a disk galaxy being ruled out. Some other galaxies are 
detected in the PC camera.\\ 

\noindent{\bf 1249+174}: This distant BL Lac ($z=0.644$) is not resolved.\\

\noindent{\bf 1255+244}: This nearby source is fully resolved in our $HST$ image.
The host galaxy is a large, circular elliptical. A disk model is ruled out.\\

\noindent{\bf 1320+084}: This source at unknown distance remains unresolved. 
The lower limit to the apparent magnitude of the host galaxy is
$m_R>22.7$~mag. Assuming the host has an absolute magnitude $M_R=-22.8$~mag, 
i.e., 1~mag fainter than the average measured for this sample, 
yields a tentative minimum distance of $z\gtrsim0.5$.\\

\noindent{\bf 1402+042}: This moderately distant source ($z=0.340$) remains unresolved. 
The radial profile is largely consistent with the PSF model and the two 
profiles become fully consistent after increasing the sky by $1\sigma$.
It is however interesting that our lower limit to the host galaxy
apparent magnitude, $m_R>19.4$~mag, is
very similar to the limit from \wsy, and is 
well within the range of luminosities observed for BL Lac hosts.
Unlike the plots for other unresolved sources, Figure~1 shows
the radial profile of the upper limit galaxy. \\ 

\noindent{\bf 1407+595}: This source was observed through the F606W filter
in order to obtain useful color information when combined with the 
F814W observation obtained by Jannuzi, Yanni \& Impey (1997).
The best-fit \dvac model gives $m_V=20.48$~mag and $r_e=1\sec .75$, 
while from the F814W observation Urry \ea (1999) derived
$m_I=18.38$~mag and $r_e=1\sec .4$. The agreement for $r_e$ is good, and the
derived color $V-I=2.1$ is consistent with the expected value
($V-I=2.6$) for an early-type galaxy at $z=0.495$.\\

\noindent{\bf 1418+54}: The source is fully resolved. The bright nuclear source
(saturated in our image) is
surrounded by a large elliptical galaxy. A smaller but sizeable galaxy, 
possibly an edge-on spiral, is visible 10.7~arcsec away. Several other 
galaxies are visible in the WF field of view.
Our best-fit \dvac model for the host gives $m_R=16.1$~mag and $r_e=3\sec .6$,
one half magnitude brighter and 50\% larger than reported
by \wsy, $m_R=16.6$~mag and $r_e=2\sec .4$\\ 

\noindent{\bf 1422+58}: This distant source ($z=0.638$, Bade \ea 1998) remains 
unresolved at $HST$ resolution.\\

\noindent{\bf 1424+24}: This bright BL Lac source remains unresolved in our
$HST$ data. The image is well exposed, enabling us to trace the radial profile
out to 7~arcsec from the nucleus, a range of $\sim 15$ magnitudes in
surface brightness. As in other similar cases, the profile of the PSF
model agrees
perfectly with the BL Lac profile over this very large range in 
brightness. Conservatively assuming the host has $M_R=-22.8$~mag, i.e., 
1~mag fainter than the average measured for this sample, our lower
limit on the host apparent magnitude ($m_R>21.0$~mag including 0.9~mag of
K correction) corresponds to a minimum distance of $z\gtrsim 0.5$.\\

\noindent{\bf 1426+42}: The object is resolved into a bright nuclear source surrounded
by an elliptical galaxy, with average radial profile 
perfectly described by a point source plus a 
\dvac model out to 7~arcsec from the center. A disk model is ruled out.\\ 

\noindent{\bf 1437+39}: This fully resolved object
is surrounded by several galaxies, suggesting it is the main 
member of a small cluster. The radial profile is well described by a point 
source plus a \dvac law. A disk model is ruled out.\\ 

\noindent{\bf 1440+122}: This BL Lac object is a member of a dumbbell system. 
At the position of the BL Lac nucleus (the eastern object in Figure~1), 
two point sources are detected, separated by
only $0''.29$. The fainter is at position angle 70$^o$.
This is an interesting gravitational lens candidate (Scarpa \ea 1999a).
The BL Lac host is an elliptical galaxy. A disk model is ruled out.\\

\noindent{\bf 1458+228}: Despite the moderately large distance ($z=0.235$) and
the short exposure time of our observation (320~s), we detected
the host galaxy of this BL Lac. The radial profile is consistent 
with either \dvac or disk models.\\

\noindent{\bf 1514-241}: AP Lib is a nearby BL Lac object for which a number of
studies have been carried out (e.g., Abraham, McHardy \& Crawford 1991). 
The $HST$ image shows a bright
point source surrounded by a large, round elliptical host.\\

\noindent{\bf 1517+656}: This source has a recently reported redshift of $z>0.7$
(Beckmann \ea, private communication) based on the presence of MgI and
FeII absorption lines. At $HST$ resolution 1517+656 shows a rather
unusual and very interesting morphology, 
with three arcs surrounding the nuclear point source, possibly tracing
an Einstein ring (Scarpa \ea 1999a). 
The average radial profile of the central object is
largely consistent with our PSF model, indicating little contribution
from a possible host galaxy.\\

\noindent{\bf 1519-273}: This source at unknown redshift remains unresolved.
We set a tentative lower limit to the distance assuming
the host has absolute magnitude $M_R=-22.8$ (i.e., 1~mag fainter than the
average measured for this sample), which gives $z\gtrsim 0.4$.\\

\noindent{\bf 1533+53}: The source is saturated and its radial
profile shows at the largest radii some deviations from the PSF
model. After increasing the sky level by $1\sigma$, 
the two profiles became consistent, so we conservatively consider
the source unresolved. 
This result is consistent with the BL Lac object being at
$z=0.89$ as tentatively reported by Bade \ea (1998).\\

\noindent{\bf 1534+014}: The BL Lac is resolved. The host is a large,
round elliptical. A disk galaxy is ruled out.
The best-fit \dvac model gives $m_R=18.2$~mag and $r_e=2\sec .0$, 
corresponding to a galaxy significantly smaller
than that reported by \wsy, $m_R=17.4$~mag and $r_e=4\sec .3$.\\

\noindent{\bf 1538+149}: For this distant BL Lac object ($z=0.605$) we have both 
a snapshot image in the F606W filter and pointed observations in F814W 
(Urry \ea 1999). In the shorter snapshot image, the object is only barely
resolved, but the host galaxy parameters are consistent with those found in the deeper
pointed observation.\\

\noindent{\bf 1544+820}: This relatively bright BL Lac at unknown redshift 
is not resolved. Using our lower limit for the host apparent magnitude,
and conservatively assuming its intrinsic luminosity is $M_R=-22.8$~mag (i.e., 
1~mag fainter than the average measured for this sample) we find 
a tentative lower limit for the distance of $z\gtrsim 0.35$.\\

\noindent{\bf 1553+113}: This well known, bright BL Lac object is heavily
saturated in our image. Thanks to the brightness of the source, we
trace its radial profile down to a surface brightness 16
magnitudes fainter than the central peak. Nonetheless, the source remains
unresolved.\\

\noindent{\bf 1704+604}: The object is resolved. The nuclear point source was
quite faint during our observation, allowing the study of the galaxy
radial profile in to the very center. The radial profile is well described
by a \dvac law, a disk model being ruled out.\\

\noindent{\bf 1722+119}: This bright BL Lac at unknown redshift remain unresolved.
We use our lower limit $m_R>21.4$~mag for the host apparent magnitude
to set a tentative lower limit to the distance of $z\gtrsim 0.55$, assuming
a host absolute magnitude of $M_R=-22.8$~mag, 
i.e., 1~mag fainter than the average measured for this sample. 
This is much larger than either the initially proposed 
redshift of $z=0.018$ (Griffiths 1989), not confirmed by Veron-Cetty \& Veron
(1993), or $z=0.159$ reported by \wsy\ (without any further reference). 
Our result suggests both values may be incorrect and we assume 
the redshift of 1722+119 remains unknown.\\

\noindent{\bf 1728+502}: This nearby ($z=0.055$) BL Lac object is fully resolved. The
host is a large, round elliptical. A disk model is ruled out.\\

\noindent{\bf 1738+47}: This source remains unresolved, with the azimuthally
averaged radial profile fully consistent with the PSF model.
We use our limit for the host apparent magnitude, $m_R>20.5$~mag,
to set a lower limit to the distance of $z>0.45$, assuming a
host galaxy absolute magnitude of $M_R=-22.8$~mag, 
i.e., 1~mag fainter than the average measured for this sample. 
This is larger than the value reported in
literature ($z=0.316$). We carefully
checked the source of this redshift and were not able to assess
its reliability. Indeed, NED reports $z=0.316$ and refers to Xu \ea (1995), 
who in turn refer to Xu \ea (1994), who
reported the source spectrum as featureless!
Therefore we conclude the redshift is unknown.\\

\noindent{\bf 1745+504}: This source is clearly resolved and several other
galaxies are present in the PC camera. 
The nuclear point source was very faint during the observation. 
A disk model for the host is ruled out.\\

\noindent{\bf 1749+701}: This bright, distant source remain unresolved.\\

\noindent{\bf 1749+096}: This moderately distant BL Lac object
is resolved into a point source surrounded by a galaxy that remains 
morphologically unclassified. The best-fit \dvac model 
gives $m_R = 18.8$~mag and $r_e=3\sec .0$, quite different from what 
was found by \wsy, $m_R = 17.7$~mag and $r_e=7\sec .1$. 
This may be due to the several stars that 
surround this low galactic latitude BL Lac object, which make it difficult
to study from the ground.\\

\noindent{\bf 1757+70}: In our $HST$ image the object is unfortunately 
located close to one edge of 
the WF4 chip. The radial profile can be extracted only 
for 180 degree and due to the presence of three stars has to be truncated
$\sim 3$~arcsec from the center. In spite of this limiting factor, and the
relatively high redshift ($z=0.407$), the source is resolved and the host 
classified as elliptical. The host has apparent magnitude $m_R=19.6$~mag, 
consistent with the value $m_R=19.5$~mag in \wsy.

\noindent{\bf 1803+784}: This moderately distant object remain unresolved.\\

\noindent{\bf 1807+698}: In our $HST$ image 3C~371 is characterized by
the presence of a jet first discovered by Nilsson \ea (1997) from the
ground. The $HST$ image was discussed by Scarpa \ea (1999b). 
The elliptical host galaxy is easily visible.\\ 

\noindent{\bf 1823+568}: For this distant source ($z=0.664$) we have a snapshot
$HST$ image in the F606W filter and a pointed $HST$ observation in the
F814W filter (Urry \ea 1999). In both cases the host galaxy is 
detected, making 1823+568 the most distant resolved BL Lac object in our sample. \\

\noindent{\bf 1853+671}: This nearby source is clearly resolved. The host
galaxy is definitely elliptical, a disk model being ruled out.\\ 

\noindent{\bf 1914-194}: The source is clearly resolved. The average radial 
profile is consistent with either \dvac and disk models. 
Few companion galaxies are detected in the PC field of view.\\ 

\noindent{\bf 1959+650}: This nearby source is fully resolved. The radial profile
is very well described by a point source plus a \dvac model.
A disk galaxy is definitely ruled out.\\ 

\noindent{\bf 2005-489}: This well studied nearby source is fully
resolved. The host galaxy is a large, round elliptical, with 
an extremely bright nucleus which dramatically impedes its
observation from the ground. The radial profile is well described by
a point source plus a \dvac model with $m_R=14.5$~mag and $r_e=5\sec .6$, 
consistent with the values found by Falomo (1996), $m_R=14.8$~mag 
and $r_e=6\sec$. A disk galaxy is ruled out.\\

\noindent{\bf 2007+777}: In our $HST$ data the source is clearly resolved.
However, the short exposure time prevents us from 
discriminating between an elliptical or a disk host.\\

\noindent{\bf 2037+521}: The source, fully resolved, is characterized by an
elliptical host galaxy. A small companion galaxy is visible 0.6~arcsec away
from the nucleus (PA$=135^\circ$), projecting well inside the host.\\

\noindent{\bf 2131-021}: This distant source ($z=1.285$) is not
resolved. No companions detected.\\

\noindent{\bf 2143+070}: The radial profile of this moderately close
BL Lac is well described by a point source plus a \dvac law. A
disk galaxy is ruled out. The best-fit parameters for the host are
$m_V=18.76$~mag and $r_e=2\sec .1$. These values agree with what was 
found in a deeper F814W observation by Jannuzi, Yanni \& Impey (1997), 
confirmed by Urry \ea (1999), $m_I=17.15$~mag and $r_e=1\sec .8$. 
The color is $V-I=1.61$, as expected for an elliptical galaxy at $z=0.237$.\\

\noindent{\bf 2149+173}: This object, for which there is no redshift
information, remains unresolved in our $HST$ data.
Conservatively assuming the host has $M_R=-22.8$~mag (i.e., 
1~mag fainter than the average measured for this sample), and 
including 0.7 magnitudes of K correction, a tentative lower
limit for the distance of $z\gtrsim 0.4$ can be set.\\

\noindent{\bf 2200+420}: The $HST$ observation of the eponymous BL Lac object
shows the source is fully resolved, albeit dominated by a very bright
nucleus. The galaxy is elliptical, as already found by 
\wsy, with apparent 
magnitude $m_R=15.5$~ and $r_e=4\sec .8$.
These values are $\sim 1$~mag fainter and a factor of 2 smaller
than found by Wurtz Stocke \& Yee (they found $r_e=10\sec .2$). 
In spite of the low redshift, these large
differences are probably due to the very bright nucleus.\\

\noindent{\bf 2201+044}: This nearby source if fully resolved. 
The host is an elliptical galaxy
and our best-fit values for the apparent magnitude and effective radius are
fully consistent with those reported by \wsy.
The most interesting result for this BL Lac is the detection of the optical
counterpart of the radio jet (see Scarpa \ea 1999b).\\ 

\noindent{\bf 2240-260}: This BL Lac object is unresolved. 
This result is consistent with the source being at $z=0.774$ (Stickel
et al. 1993), based on the detection of two very weak emission lines
that have never been confirmed.\\

\noindent{\bf 2254+074}: We have both pointed (F814W) and snapshot
(F606W) observations of this BL Lac object. 
The source is fully resolved and classified as elliptical 
in both cases. The best-fit parameters are $m_V=17.4$~mag and $r_e=4\sec .9$
for the F606W observation,
and $m_I=15.9$~mag and $r_e=4\sec .8$ for the F814W observation.
The derived color is $V-I=1.5$, as expected for an early-type 
galaxy at $z=0.190$.\\

\noindent{\bf 2326+174}: This BL Lac object has a round elliptical host galaxy.
A disk model for the host is ruled out.\\

\noindent{\bf 2344+514}: A large elliptical host galaxy dominates the $HST$ image of
this nearby BL Lac object. Its radial profile follows
particularly well a \dvac law.\\

\noindent{\bf 2356-309}: This nearby source is fully resolved. The host is
an elliptical, with a radial profile well fitted by a \dvac law.
A disk model is ruled out. Our values for the host galaxy magnitude
and effective radius, $m_R=17.21$~mag and $r_e=1\sec .85$, 
agree well with the values reported by Falomo \ea (1991),
$m_R=17.24$~mag and $r_e=3\sec .5$.\\


\newpage


\begin{figure}
   \vspace{-3cm}
  \centerline{
    \psfig{file=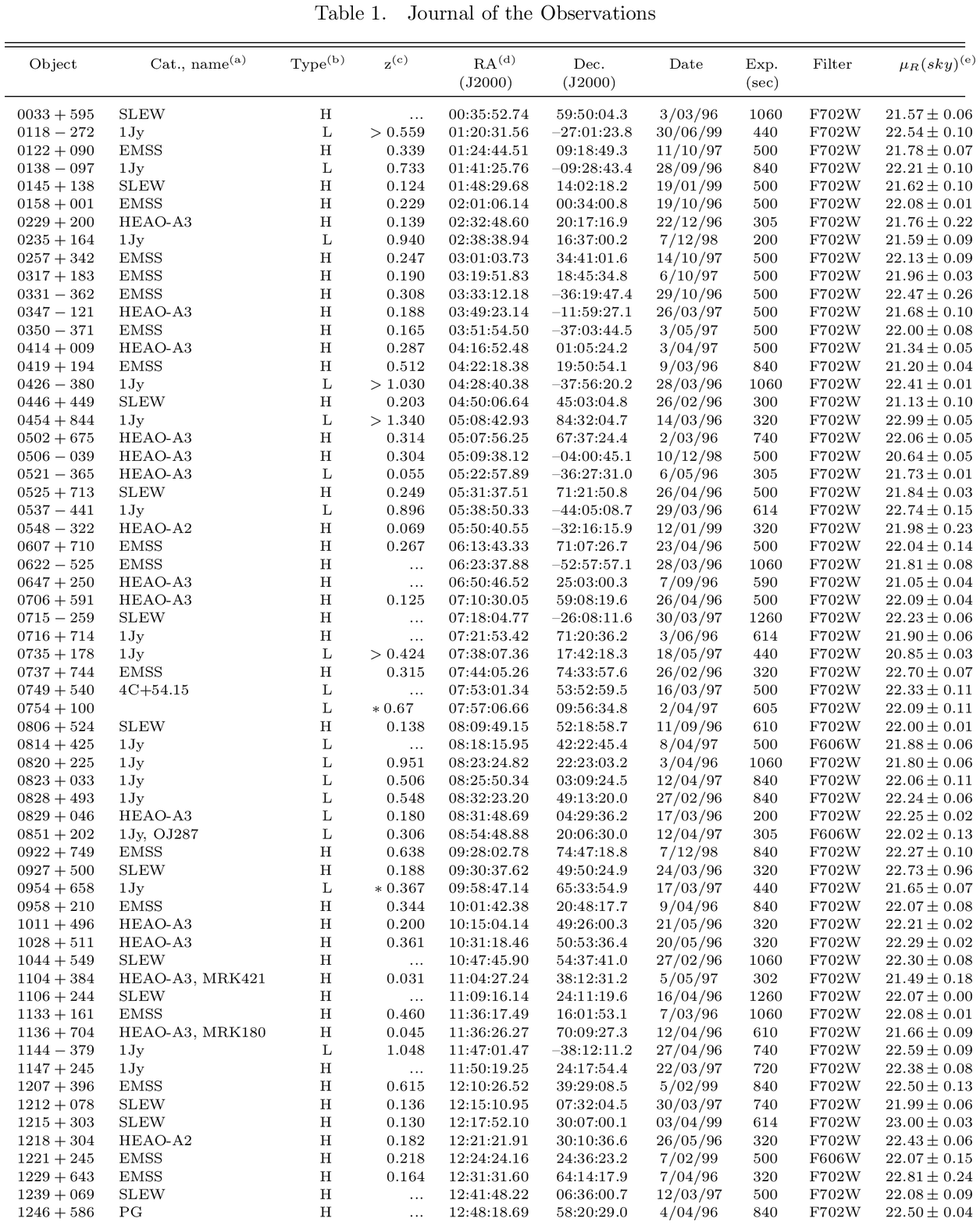}
  }
\end{figure}

\newpage
\begin{figure}
   \vspace{-3cm}
  \centerline{
    \psfig{file=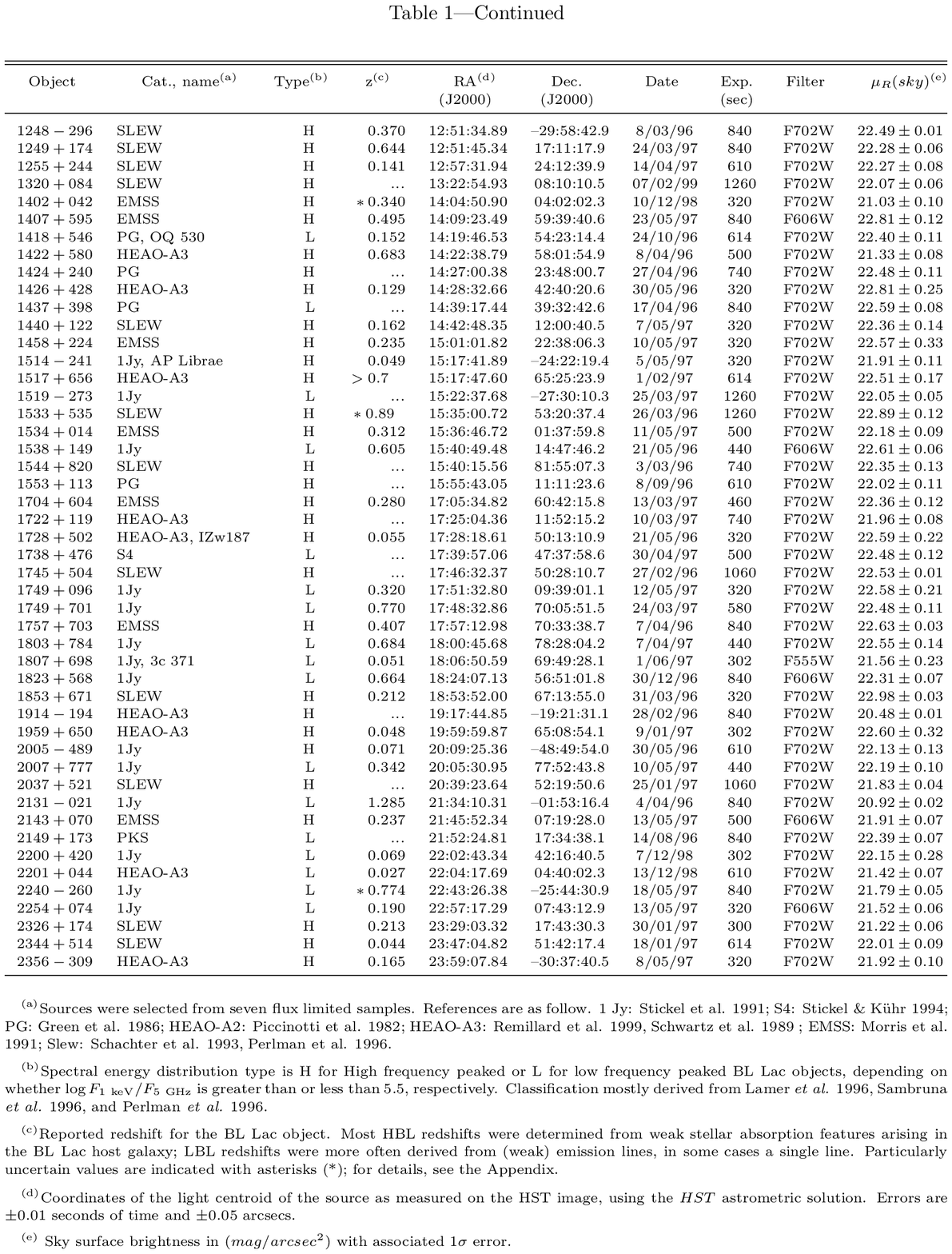}
  }
\end{figure}

\newpage

\begin{deluxetable}{rrr}
\small
\tablenum{2}
\tablewidth{2.5in}
\tablecaption{Scattered Light Coefficients}
\tablehead{ 
\colhead{Filter} & \colhead{A} & \colhead{B}
}
\startdata
F814W  &$ 5.6 \times 10^{-4} $&$ -0.42 $\nl
F702W  &$ 7.3 \times 10^{-4} $&$ -0.50 $\nl
F606W  &$ 1.3 \times 10^{-3} $&$ -0.48 $\nl
F555W  &$ 8.0 \times 10^{-4} $&$ -0.48 $\nl
\enddata
\end{deluxetable}


\begin{figure}
   \vspace{-2.5cm}
  \centerline{
    \psfig{file=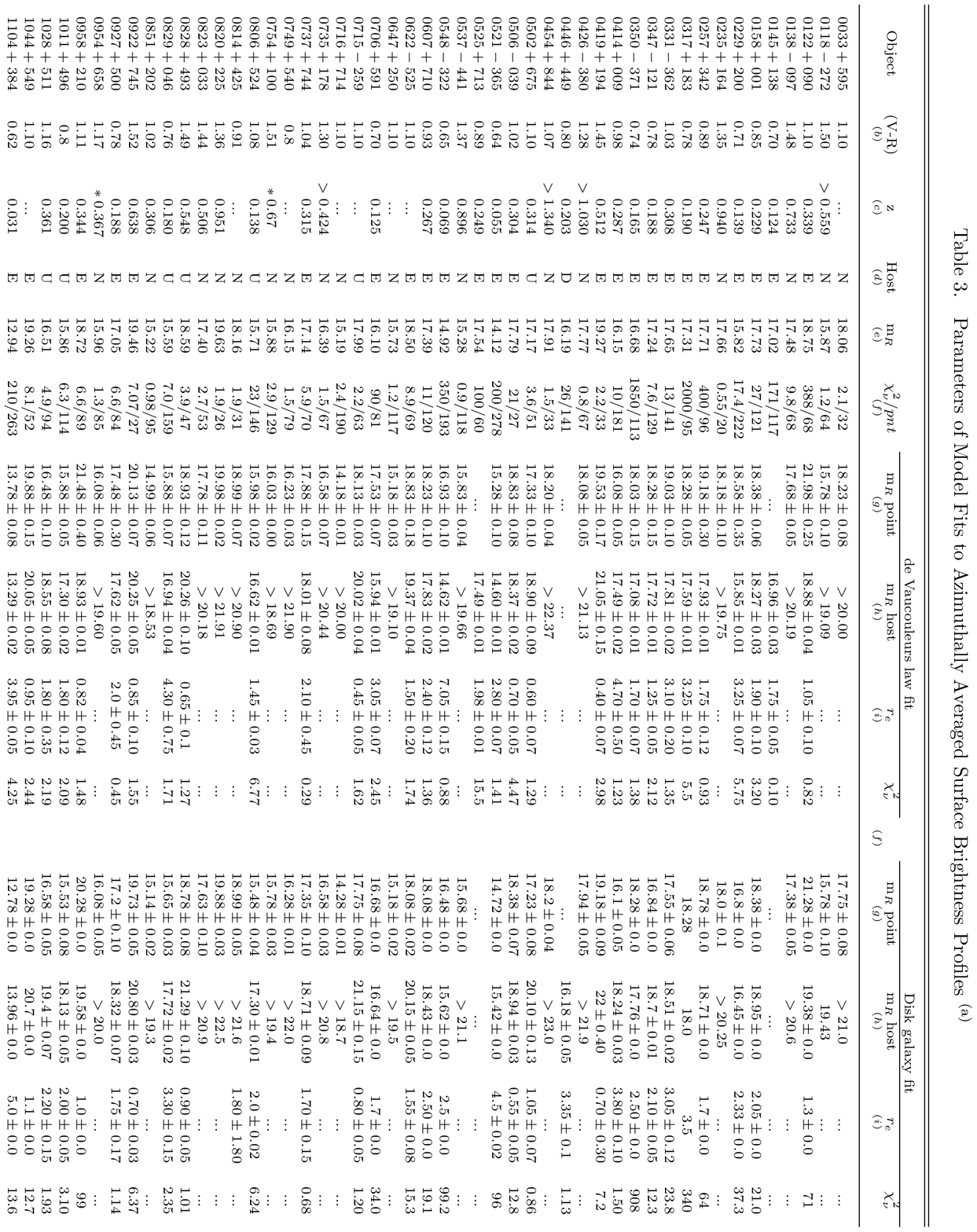}
  }
\end{figure}

\begin{figure}
   \vspace{-2.5cm}
  \centerline{
    \psfig{file=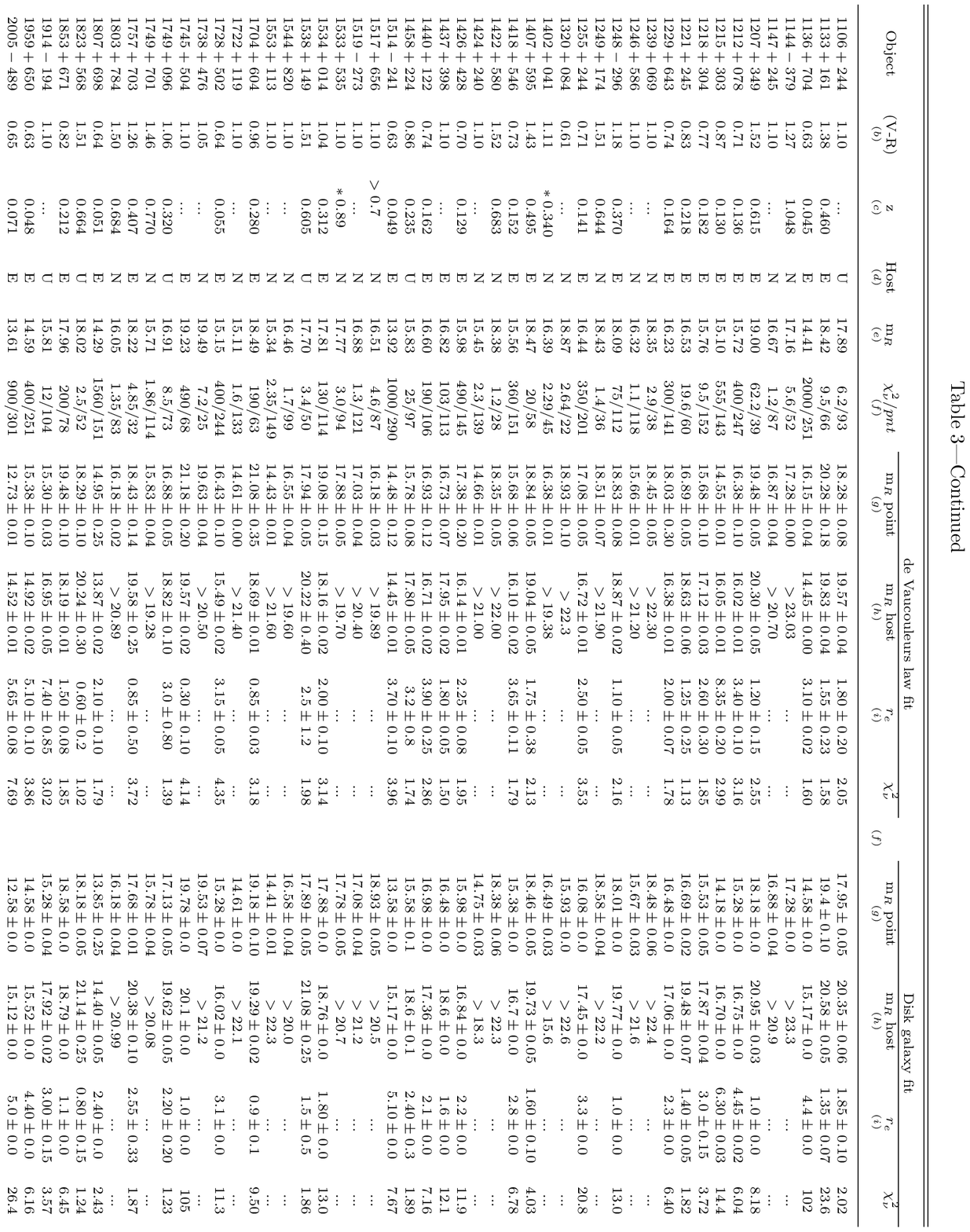}
  }
\end{figure}

\begin{figure}
   \vspace{-2.5cm}
  \centerline{
    \psfig{file=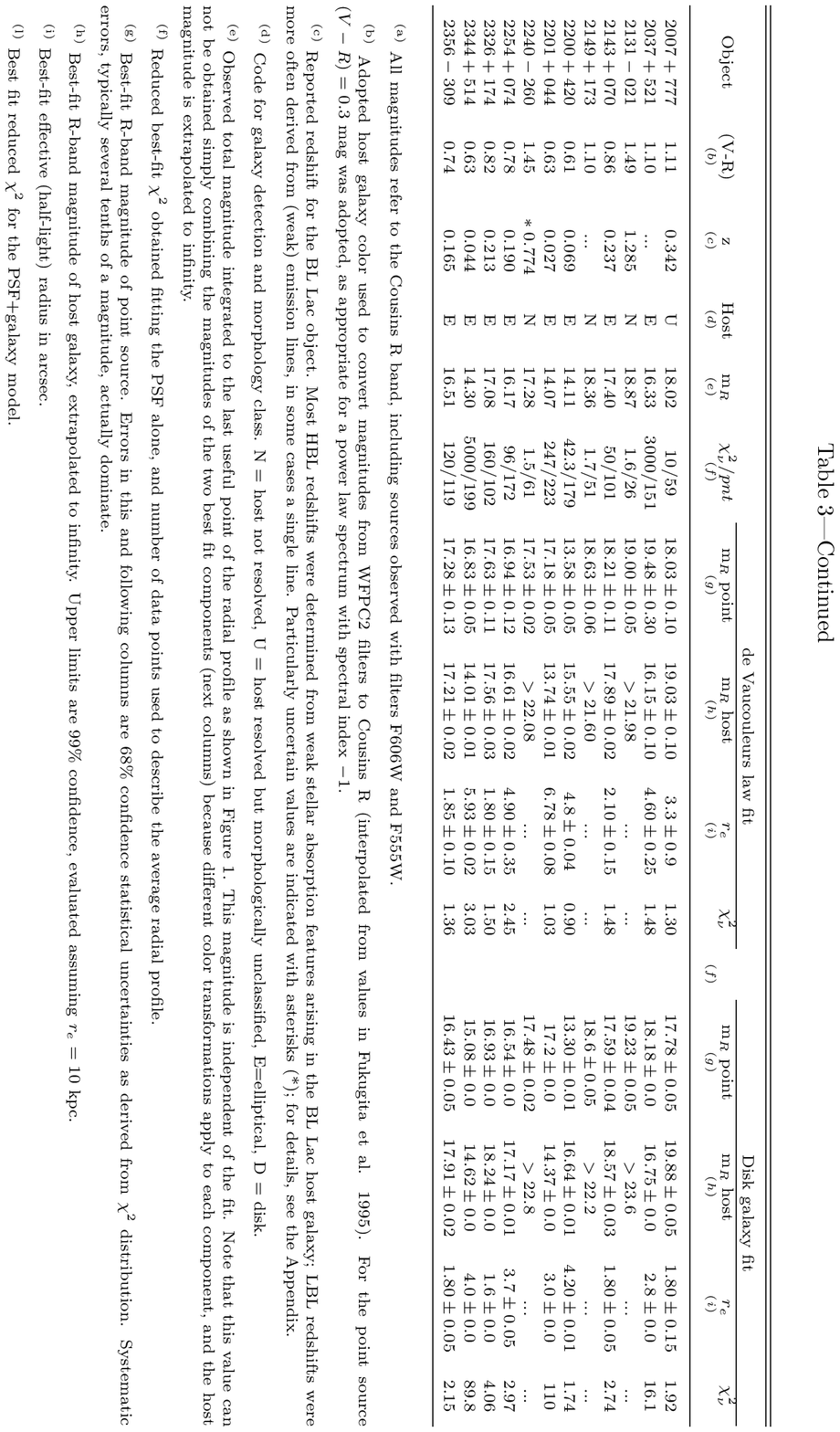}
  }
\end{figure}

\newpage

\begin{figure}
\caption{ {\bf Figure 1.}
For each BL Lac object up to four panels are
shown. The top panel is the final combined, cosmic-ray-cleaned image
in gray scale. The next panel is a contour plot of the same image,
which better reveals a large range in surface brightness. Contours
start at the surface brightness (in mag~arcsec$^{-2}$) indicated at
the bottom of each figure and are separated by 0.5 magnitudes. The
two upper panels have the same size and orientation, which are
reported on the contour panel; north and east are marked. The third
panel shows the average surface brightness radial profile ({\it filled
squares}), and the best-fit PSF plus \dvac model (or only the PSF for
unresolved sources). {\it Dotted line:} the PSF model. {\it Dashed
line:} \dvac model convolved with the PSF. {\it Solid line:} sum of
the two components. When not visible, the dashed line is superimposed
on the solid line.
In several images the
central point source is saturated (sometime heavily). Saturation
shows up usually in the first 2 or 3 pixels of the radial profile,
either as a flat plateau (e.g., 1914-194), or as a region of decreasing
flux (e.g., 1144-379). In some other cases the saturation was not
that heavy and the software was able to recover partially the true
nuclear brightness by combining long and short (unsaturated) images; 
the radial profile then shows a modest, less-than-proportional increase of flux
going toward the center (e.g., 1044+549 and 1407+595). For comparison, 
two cases of unsaturated radial profiles are 1426+428 and 1440+122.
For the three exceptional objects 0145+138, 0446+449, and 0525+713,
which have no detectable point source, we fitted the radial profile 
with a galaxy model only. For 0446+449 the profile is a perfect
exponential, the only case of a disk galaxy in the entire sample (and
one with no bright nucleus).
For resolved BL Lacs, a fourth panel shows the $\chi^2$ contours for
the two parameters of interest in the PSF-plus-\dvac fit, $r_e$ and
$m_R$, with best-fit values marked by a square. The contours
represent 68\%, 95\%, 99\% probabilities.
}
\end{figure}

\begin{figure}
\psfig{file=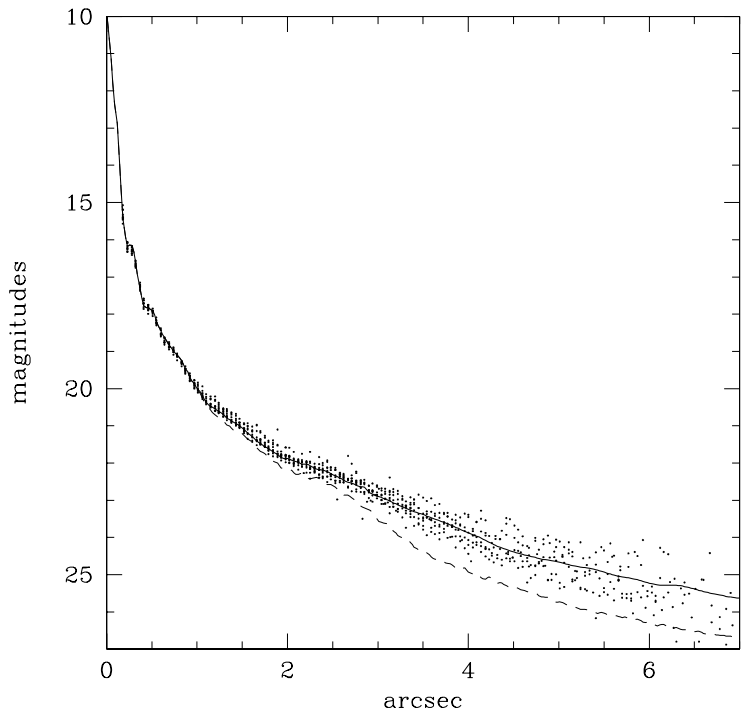,width=12cm}
\caption{
{\bf Figure 2.} Radial profile derived from several bright
stars (dots), compared with the Tiny Tim-generated PSF ({\it dashed line})
and our composite PSF model ({\it solid line}).
}
\end{figure}

\begin{figure}
\psfig{file=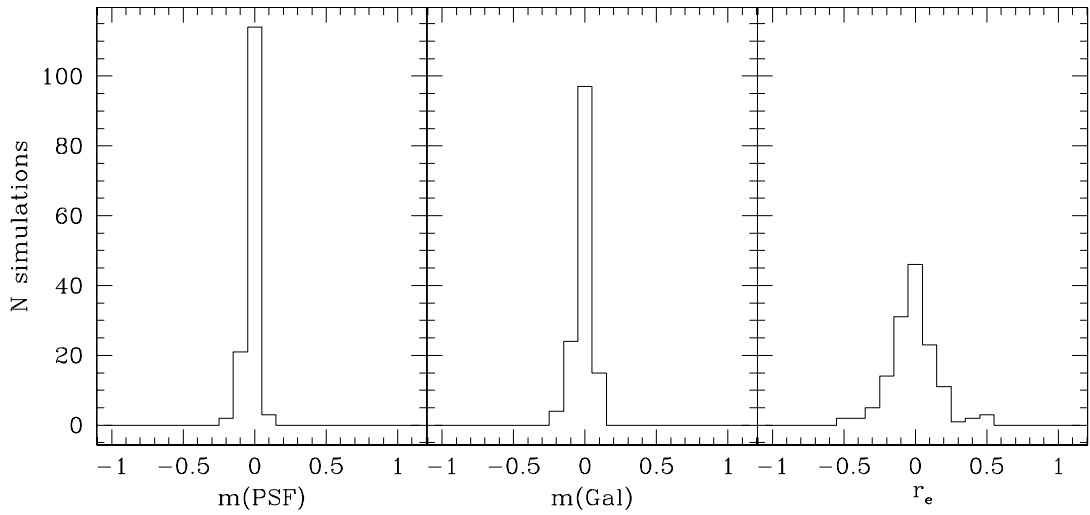,width=16cm}
\caption{
{\bf Figure 3.}
The distribution of best-fit minus input values of nuclear apparent
magnitude, host galaxy apparent magnitude, and effective radius (in
arcseconds) for simulated $HST$ images, analyzed according to our
standard procedures. The mean values are centered at zero and the
dispersions are within our estimated statistical errors, 
meaning we recover the
true value with no systematic deviations and with high accuracy.
In the simulations, nuclear magnitudes were varied from 15 to 18 mag, the host
magnitudes from 15 to 21 mag, and the effective radii from 0.8 to 5 arcsec,
which represent the values most often observed in our data.
}
\end{figure}

\begin{figure}
\psfig{file=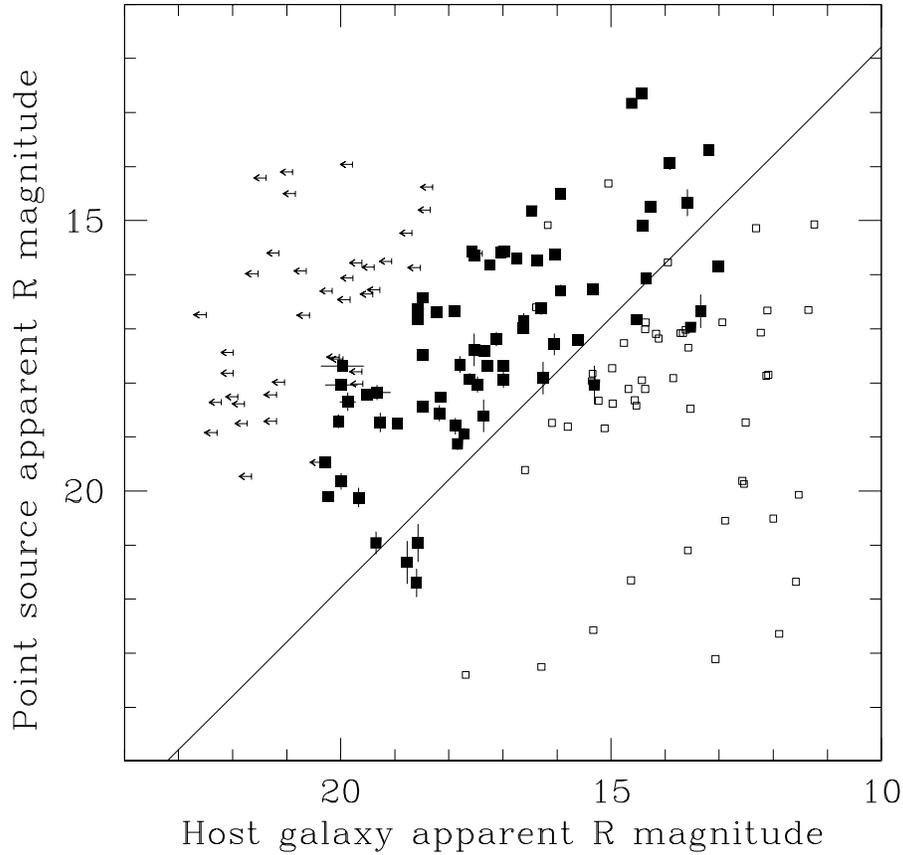,width=15cm}
\caption{
{\bf Figure 4.} Host galaxy versus nuclear magnitude for all
\ntot BL Lac objects ({\it solid squares represent resolved sources, 
while arrows are upper limits for unresolved objects}). The solid line is
the locus for an elliptical galaxy with a nuclear luminosity
that reduces an intrinsic 50\% 4000\AA\ break to 25\%, the usual
distinction between a BL Lac object and a radio galaxy. 
The nuclear non-thermal component is assumed to have spectral index 
$\alpha = -1.5$, which includes all but the steepest BL Lacs.
Objects with flatter spectral indices would be located to the left
of this line. Objects above the line have bright
nuclei compared to their host galaxies and are classified as BL Lacs. 
Below the line the light from the galaxy
dominates and the source is classified as radio galaxy.
Sources can cross the limit because of nuclear
flux variability. Radio galaxies ({\it open squares};
Govoni \ea 1999, Chiaberge \ea 1999) smoothly fill the lower half
of the diagram, showing that the distinction from BL Lac objects occurs
at an arbitrary point.
}
\end{figure}

\begin{figure}
\psfig{file=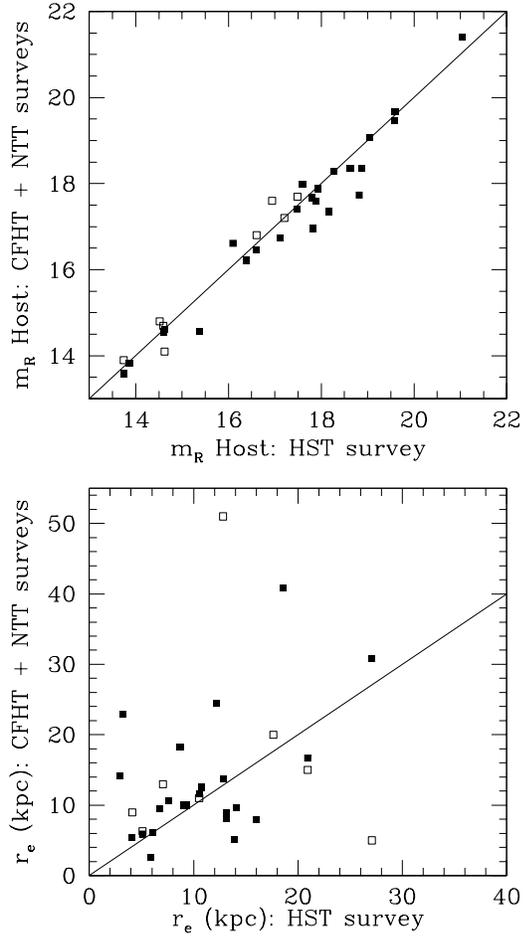,width=13cm}
\caption{
{\bf Figure 5.} Comparison of our $HST$ results with those from
two large ground-based surveys. {\it Filled squares:} 
Canada-France-Hawaii-Telescope survey (Wurtz, Stocke \& Yee 1996).
{\it Open squares:} ESO-NTT survey  (Falomo 1996).
The magnitudes agree well, with an average difference of 0.1~mag
and an rms dispersion of 0.4~mag. The scatter in measured radius
is much larger.
}
\end{figure}

\begin{figure}
\psfig{file=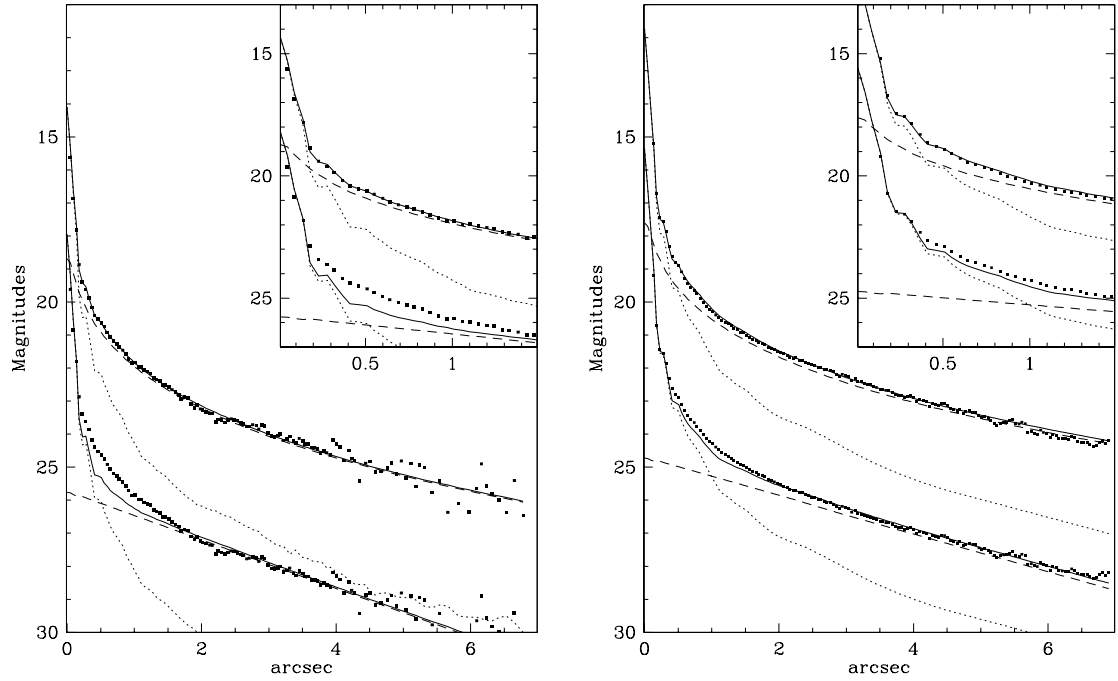,width=15cm}
\caption{
{\bf Figure 6.} Comparison of the best \dvac and disk fits for 
two illustrative sources, 0607+710 ({\it left}) and 1418+546 ({\it right}).
To compare the two, the disk fit is shifted downward by 4~mag.
The inset shows an enlargement of the central portion of the radial profile. 
The PSF is shown as a dotted line, the galaxy as a dashed line, and the
sum of the two components as a solid line.
In both cases, the exponential disk fails by a small but highly significant
amount to reproduce the observed light in the central 1.5~arcsec, which is
the case for most of the BL Lac host galaxies. 
}
\end{figure}

\begin{figure}
\psfig{file=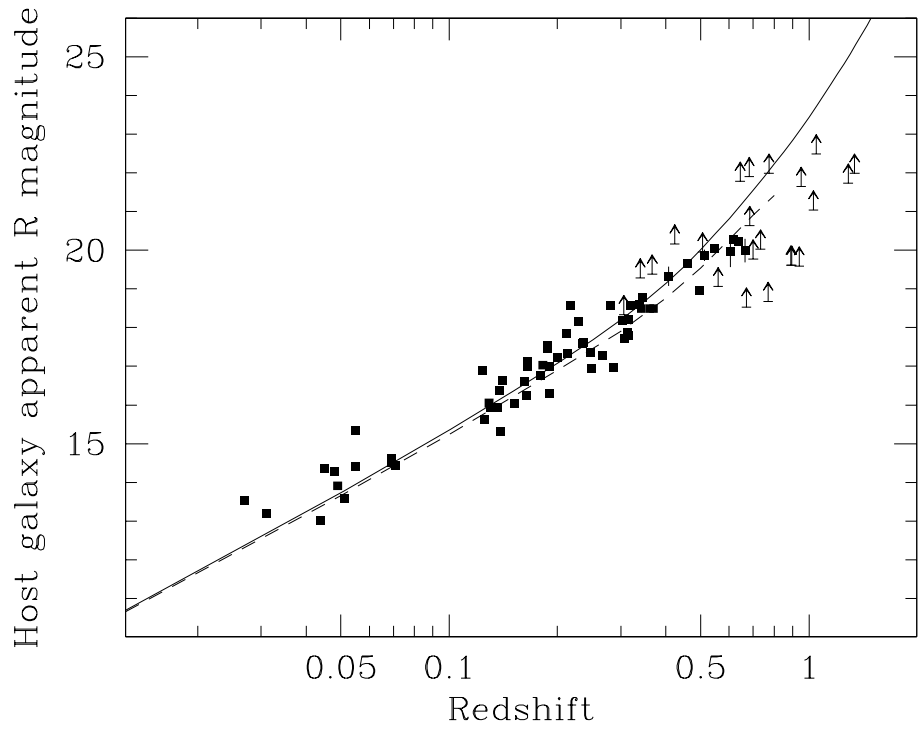,width=15cm}
\caption{
{\bf Figure 7.} Hubble diagram for BL Lac host galaxies. 
The solid line is for an object of absolute magnitude $M_R=-23.7$~mag, 
while the dashed line is for the same object including passive evolution 
(Bressan \ea 1994). The data ({\it filled squares}) are
consistent with either line, with few resolved host galaxies 
at $z\gtrsim 0.6$.
}
\end{figure}

\begin{figure}
\psfig{file=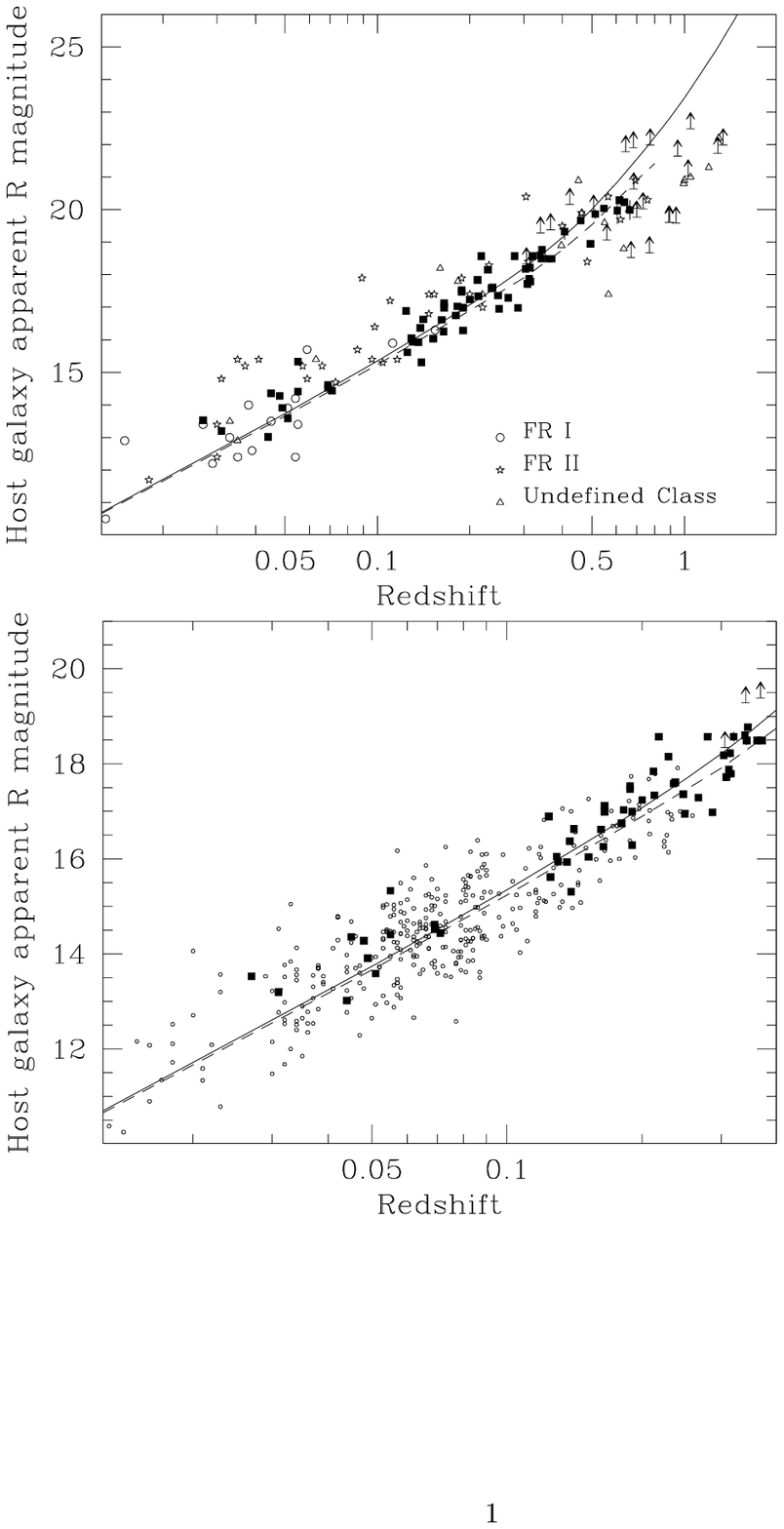,width=12cm}
\caption{
{\bf Figure 8.} Hubble diagram for BL Lac host galaxies 
and radio galaxies.
\noindent{\it Upper panel}: BL Lacs ({\it solid squares}) are compared with 
radio galaxies from the 2~Jy sample ({\it open symbols}; Morganti \ea 1993). 
\noindent{\it Lower panel}: The low-redshift region from the previous panel,
with radio galaxy data ({\it dots}) from Ledlow \& Owen (1995) and 
Govoni \ea (1999).
}
\end{figure}

\begin{figure}
\psfig{file=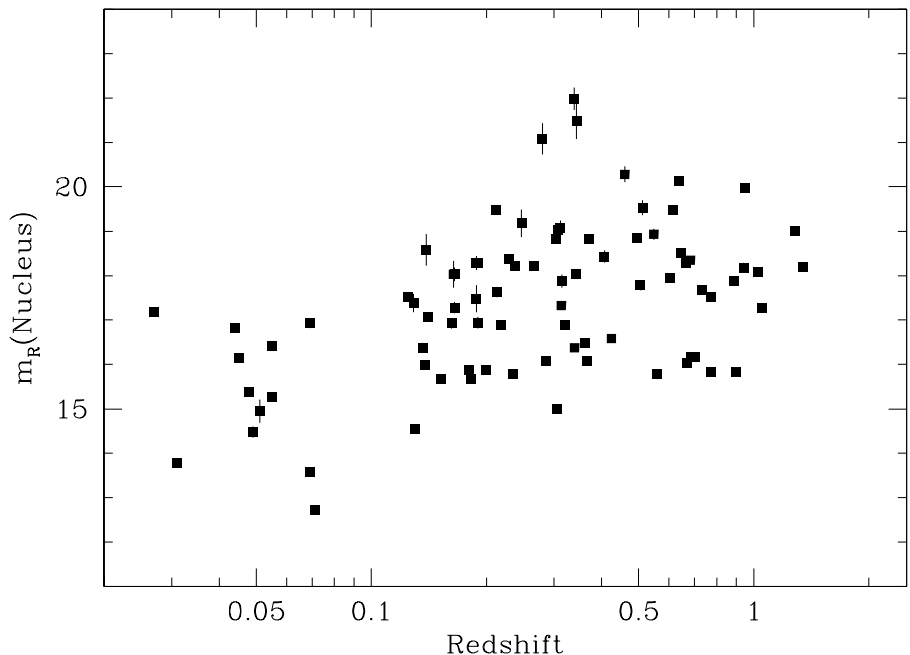,width=14cm}
\caption{
{\bf Figure 9.} Apparent R magnitude of the nuclear component,
as a function of redshift. All sources except
0145+138, 0446+449, and 0525+713, which have no central point sources,
are shown. For resolved 
sources the best-fit point source magnitude is plotted, as derived from
a PSF plus \dvac model even when the host remains morphologically
unclassified. 
Few of the nuclei are fainter than $m_R\sim20$~mag because of the effective
flux limit imposed by optical identification of radio or X-ray samples.
In strong contrast to the Hubble diagram for the host galaxies,
the apparent magnitude of the nuclear component does not vary strongly
with distance, as expected in flux-limited samples, for which luminosity 
increases with redshift. 
The shallow envelope in the lower right reflects the luminosity function, 
while the upper left would be populated by radio galaxies.
}
\end{figure}

\begin{figure}
\psfig{file=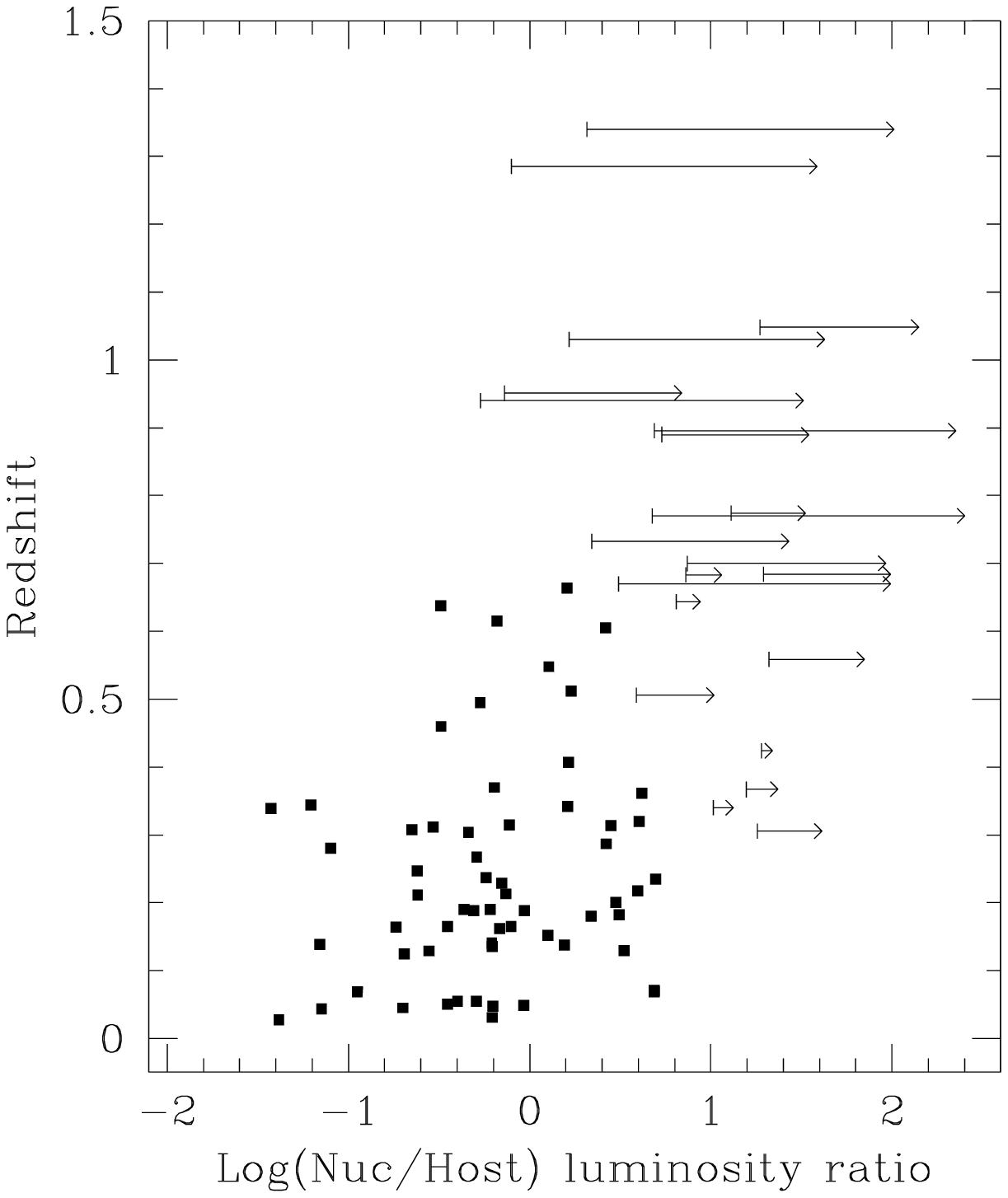,width=15cm}
\caption{
{\bf Figure 10.} The $L_{nuc}/L_{host}$ luminosity ratio as a
function of distance, using K-corrected data.
For unresolved objects, we plot a range of ratios ({\it horizontal arrows}). 
The minimum ratio (leftmost extreme of the arrow) 
corresponds to the upper limit to the host galaxy magnitude.
The maximum ratio (rightmost extreme) is the ratio assuming a galaxy 
luminosity $M_R=-22.8$~mag (1~mag fainter than the average of the sample).
The ratio of nucleus to host galaxy luminosity increases with redshift,
as the point source luminosity increases.
}
\end{figure}


\noindent
\begin{references}

\reference{} Abraham R.G., McHardy I.M. \& Crawford C.S. 1991, MNRAS 252, 482
\reference{} Bade N., Beckmann V., Douglas N.G., \ea 1998, A\&A 334, 459
\reference{} Bahcall J.N., Kirhakos S., Saxe D.H. \& Schneider 1997, ApJ 479, 642
\reference{} Bressan A., Chiosi C. \& Fagotto F. 1994, ApJS 94, 63
\reference{} Browne I.W.A. 1989, in BL Lac Objects, ed. L. Maraschi, 
	T. Maccaro, \& M. H. Ulrich, p 401
\reference{} Carswell R.F., Strittmatter P.A., Williams R.E., Kinman T.D.
	\& Serkowski K. 1974, ApJ 190, L101
\reference{} Chiaberge M, Capetti A. \& Celotti A. 1999, A\&A, in press
\reference{} Disney M.J., Boyce P.J., Blades J.C. \ea 1995 Nature 376, 150
\reference{} Dressler A. \& Shectman S.A. 1987, AJ 94, 899
\reference{} Dunlop J.S., Peacock J.A., Savage A., Lilly S.J., Heasley J.N.,
	Simon A.J.B. 1989, MNRAS 238, 1171
\reference{} Fanaroff B. \& Riley J.M. 1974, MNRAS 167, 31P
\reference{} Falomo R. 1991, AJ 101, 821
\reference{} Falomo R. 1996, MNRAS 283, 241
\reference{} Falomo R., Kotilainen J., Pursimo T., Sillanp\"a\"a A., 
	Takalo L. \& Heidt J. 1997, A\&A 321, 374
\reference{} Falomo R., Pesce J.E., \& Treves A. 1993 ApJ 411, L63
\reference{} Falomo R., Pesce J.E., \& Treves A. 1995 ApJ 438, L9
\reference{} Falomo R., Scarpa R., \& Bersanelli M. 1994, ApJS 93, 125
\reference{} Falomo R., Urry C.M., Pesce J.E., Scarpa R., 
	Giavalisco M. \& Treves A. 1997, ApJ 476, 113
\reference{} Falomo R., Urry C.M., Pesce J.E., Scarpa R., 
	\& Treves A. 2000 in preparation
\reference{} Fichtel C.E., Bertsch D.L., Chiang J. \ea 1994 ApJS 94, 551
\reference{} Fukugita M, Shimasaku K. \& Ichikawa T. 1995, PASP 107, 945
\reference{} Govoni F., Falomo R., Fasano G. \& Scarpa R. 1999, A\&A in press
\reference{} Green R.F., Schmidt M. \& Liebert J. 1986, ApJS 61, 305
\reference{} Griffiths R.E., Wilson A.S., Ward M.J., Tapia S. \& 
	Ulvestad J.S. 1989, MNRAS 240, 33
\reference{} Halpern J.P., Impey C.D., Bothun G.D., Tapia S., Skillman E.D.
	Wilson A.S. \& Meurs E.J.A. 1986 ApJ 302, 71
\reference{} Holtzman, J. A., Burrows, C. J., Casertano, S., Hester, J. J.,
	Trauger, J. T., Watson, A. M., \& Worthey, G. 1995, PASP, 107, 1065
\reference{} Hooper E.J., Impey C.D., \& Foltz C.B. 1997, ApJ, 480, L95
\reference{} Jannuzi B.T., Yanny B. \& Impey C. 1997, ApJ 491, 146
\reference{} Kollgaard R.I., Palma C., Laurent-Muehleisen S.A. \& 
	Feigelson E.D. 1996, ApJ 465, 115
\reference{} Kollgaard R. I., Wardle J. F. C., Roberts D. H., \& 
	Gabuzda D. C. 1992, AJ 104, 1687
\reference{} Kotilainen J.K., Falomo R. \& Scarpa R. 1998, A\&A 336, 479
\reference{} Krist, J. 1995, in Astronomical Data Analysis, Software
	and Systems IV, eds. R. Shaw et al., 
	(San Francisco: Astr. Soc. Pac.), p. 349
\reference{} Laing R.A. 1994, in {\it The Physics of Active Galaxies}. 
	ASP Conference Series, G.V. Bicknell, M.A. Dopita, and 
	P.J. Quinn, Eds., Vol. 54, p. 227
\reference{} Lamer G., Brunner H. \& Staubert R. 1996, A\&A 311, 384
\reference{} Lawrence C.R., Pearson T.J., Readhead A.C.S. \& 
	Unwin S.C. 1986 AJ 91, 494
\reference{} Lawrence C.R., Zucker J.R.,Readhead A.C.S.,Unwin S. C., 
	Pearson T.J., \& Xu W. 1996 ApJS 107, 541
\reference{} Ledlow M.J. \& Owen F.N. 1995, AJ 109, 853
\reference{} Malkan M.A., Gorjian V. \& Tam R. 1998, ApJS 117, 25
\reference{} McHardy I.M., Abraham R.J., Crawford C.S., Ulrich M.-H.,
	Mock P.C. \& Vanderspek R.K. 1991, MNRAS, 249, 742
\reference{} McHardy I.M., Luppino G.A., George I.M., Abraham R.J., 
	Cooke B.A. 1992, MNRAS 256, 655
\reference{} McHardy I.M., Merrifield M.R., Abraham R.J., \& Crawford C.S.
	1994, MNRAS 268, 681
\reference{} McLeod K.K., Rieke J.H. \& Storrie-Lombardi L.J. 1999, 
	ApJ 511, L67
\reference{} McLure R.J., Kukula M.J., Dunlop J.S., Baum S.A., O'Dea C.P. \&
	Hughes D.H. 1999, MNRAS in press
\reference{} Morganti R., Killen N.E.B. \& Tadhunter C.N. 1993, MNRAS 263, 1023
\reference{} Morris S.L., Stocke J.T., Gioia I.M., Schild R.E., 
	Wolter A., Maccacaro T. \& Della Ceca R. 1991, ApJ 380, 49
\reference{} Nilsson K., Heidt J., Pursimo T., Sillanp\"a\"a A., 
	Takalo L.O. \& J\"ager 1997 ApJ 484, L107
\reference{} Owen F.N., Ledlow M.J. \& Keel W.C. 1996, AJ 111, 53
\reference{} Perez-Fournon I \& Biermann P. 1984, A\&A 130, L13
\reference{} Perlman E.S., Stocke J.T., Wang Q.D. \& Morris S.L. 1996, 
	ApJ 456, 451
\reference{} Perlman E.S., Stocke J.T., Shaffer D.B., Carilli C.L. \&
	MA C. 1994 ApJ 424, 69
\reference{} Persic M. \& Salucci P. 1986, in {\it Structure and Evolution 
	of Active Galactic Nuclei}, ed. G. Giuricin, F. Mardirossian, 
	M. Mezzetti and	M. Ramella, p. 657
\reference{} Pesce J.E., Falomo R. \& Treves A. 1995, AJ 110, 1554
\reference{} Piccinotti G., Mushotzky R.F., Boldt E.A., Holt S.S., 
	Marshall F.E., Serlemitsos P.J. \& Shafer R.A. 1982 ApJ 253, 485
\reference{} Polomski E., Vennes S., Thorstensen J.R., Mathioudakis M. \&
	Falco E.E. 1997 ApJ 486, 179
\reference{} Remillard R. \ea 1999, in preparation
\reference{} Romanishin W. 1992, ApJ 401, L65
\reference{} Sambruna R., Maraschi L. \& Urry C.M. 1996, ApJ 463, 444
\reference{} Scarpa R. \& Falomo R. 1997, A\&A 325, 109
\reference{} Scarpa R., Urry C.M., Falomo R., Pesce J.E., Webster R., O'Dowd M.
	\& Treves A. 1999a, ApJ 521, in press
\reference{} Scarpa R., Urry C.M., Falomo R. \& Treves A. 1999b, ApJ 526, 
	in press
\reference{} Schachter J.F., Stocke J.T., Perlman E., \ea 
	1993, ApJ 412, 541
\reference{} Schwartz, D. A., Brissenden, R. J. V., Tuohy, RT. R., 
	Feigelson, E. D., Hertz, P. L., \& Remillard, R. A., 1989, 
	in {\it BL Lac Objects}, ed. L. Maraschi, T. Maccacaro, M.-H. Ulrich 
	(Berlin: Springer-Verlag), p. 208
\reference{} Schwartz D.A. \& Ku W.H.M. 1983, ApJ 266, 459
\reference{} Sillanp\"a\"a A., Takalo L.O., Nilsson K., Pursimo T. \&
	Pietil\"a 1998, in {\it BL Lac Phenomenon}, A.S.P. Conf. Ser., 
	Vol. 159, ed. L.T. Takalo and A. Sillanp\"a\"a, p. 395
\reference{} Stickel M., Fried J.W., \& K\"uhr H. 1993, A\&AS 98, 393
\reference{} Stickel M. \& K\"uhr H. 1994 A\&AS, 103, 349
\reference{} Stickel M., Meisenheimer K. \& K\"uhr H. 1994, A\&AS 105, 211
\reference{} Stickel M., Padovani P., Urry C.M., Fried J.W., \& Kuhr H. 
	1991, ApJ 374, 431
\reference{} Stocke J.T., Morris S.L., Gioia I.M., Maccacaro T., Schild R.,
	Wolter A., Fleming T.A. \& Henry J.P. 1991, ApJS 76, 813
\reference{} Stoke J.T. \& Rector T.A. 1997, ApJ 489, L17
\reference{} Stocke J.T., Wurtz R., Wang Q., Elston R. \& Jannuzi B.T. 1992
	ApJ 400, L17
\reference{} Stocke J.T., Wurtz R.E. \& Perlman E.S. 1995, ApJ 454, 55
\reference{} Ulrich, M.H. 1989, in BL Lac Objects, ed. L. Maraschi, 
	T. Maccacaro, \& M. H. Ulrich, p 45
\reference{} Urry C. M. \& Padovani P. 1995, PASP 107, 803
\reference{} Urry C. M., Falomo R., Scarpa R., Pesce J. E., Treves A. \& 
	Giavalisco M. 1999, ApJ 512, 88
\reference{} Urry C. M., Scarpa R., O'Dowd M., Falomo R., Giavalisco M. 
	Pesce J. E., Treves A. \& 2000, ApJ submitted (Paper II)
\reference{} V\'eron-Cetty M. P. \& V\'eron P. 1993, A\&AS 100, 521
\reference{} Wall J.V. \& Peacock J.A. 1985, MNRAS 216, 173
\reference{} Wills D. \& Wills B.J. 1976, ApJS 31, 143
\reference{} Wisniewski W.Z., Sitko M.L. \& Sitko A.K. 1986 MNRAS 219, 299
\reference{} Wurtz R., Stoke J.T. \& Yee H.K.C. 1996, ApJS 103, 109
\reference{} Yanny B., Jannuzi B.T. \& Impey C. 1997, ApJ 484, L113
\reference{} Xu W., Lawrence C.R., Readhead A.C.S. \& Pearson T.J. 
	1994, AJ 108, 395
\reference{} Xu W., Readhead A.C.S., Pearson T.J., Polatidis A.G. \& 
	Wilkinson P.N. 1995, ApJS 99, 297

\end{references}
\end{document}